\pdfoutput=1
\documentclass[10pt,a4paper]{article}

\PassOptionsToPackage{hyphens}{url}
\usepackage[top=1.0in,bottom=1.1in,left=1.1in,right=1.1in]{geometry}
\usepackage{amsmath,amssymb,amsthm,mathtools}
\usepackage{booktabs}
\usepackage{pifont}
\usepackage{float}
\usepackage{longtable}
\usepackage{pdflscape}
\usepackage{tabularx}
\usepackage{caption}
\usepackage{ragged2e}
\newcolumntype{Y}{>{\raggedright\arraybackslash}X}
\newcolumntype{P}[1]{>{\raggedright\arraybackslash}p{#1}}

\usepackage{tikz}
\usetikzlibrary{positioning,arrows.meta}
\usepackage{hyperref}
\usepackage{url}
\usepackage[final,expansion=false]{microtype}
\newcommand{\appendixlayout}{\raggedbottom}

\makeatletter
\g@addto@macro\UrlBreaks{%
  \do\_\do\/\do\.\do\-\do\:\do\,\do\;%
  \do\A\do\B\do\C\do\D\do\E\do\F\do\G\do\H\do\I\do\J\do\K\do\L\do\M%
\do\N\do\O\do\P\do\Q\do\R\do\S\do\T\do\U\do\V\do\W\do\X\do\Y\do\Z}
\g@addto@macro\UrlBreaks{%
  \do\a\do\b\do\c\do\d\do\e\do\f\do\g\do\h\do\i\do\j\do\k\do\l\do\m%
\do\n\do\o\do\p\do\q\do\r\do\s\do\t\do\u\do\v\do\w\do\x\do\y\do\z}
\makeatother
\newtheorem{theorem}{Theorem}[section]

\newtheorem{corollary}[theorem]{Corollary}
\newtheorem{proposition}[theorem]{Proposition}
\newtheorem{definition}[theorem]{Definition}

\newtheorem{remark}[theorem]{Remark}

\title{The Orientation Boundary:\\
A Machine-Checked Structural Theory of Termination Proofs for Step-Duplicating Recursion}

\author{Moses Rahnama\\Mina Analytics}

\date{March 2026}

\begin{document}

\maketitle

\begin{abstract}
  The rewrite rule
  \[
    \mathrm{recur}(b,s,\mathrm{succ}(n))\to
    \mathrm{wrap}(s,\mathrm{recur}(b,s,n))
  \]
  copies its step argument, the payload, and consumes one recursion counter. It terminates; we ask which direct measures prove it. For measures built from counter and payload by constants, sums, products, maxima, and natural scalar multiples, those that orient every root successor step of the free recursor are those that ignore the payload and increase strictly with the counter. This set is the Orientation Boundary. The mechanism compares the cost of the copied payload with the gain per counter step: every natural-valued measure of any shape (linear, polynomial, matrix, or other) that keeps both wrapper arguments with bounded counter gain fails, and twelve direct measure families fail by this comparison under their hypotheses. Termination is proved by violating a barrier hypothesis or using information outside the direct classes: dependency-pair projection, nonlinear counter-payload coupling, or a path order. Seventy-six termination methods from the literature receive a machine-checked termination verdict on the free recursor. For constructor-local natural polynomial interpretations of any degree, orientation of the two free-recursor rules is decidable at unit successor slope, and at higher slopes its undecided remainder contains the open zero problem for two-variable integer polynomials over $\mathbb N^2$. The free recursor is strongly normalizing and confluent under all contexts, so orientation separates from termination. Every numbered result is machine-checked in Lean~4; TTT2 and CeTA certify the free recursor and a fixed system containing it, and the barrier theorems predict the MAYBE answers of TTT2's Knuth-Bendix strategy there, and of its matrix strategies when both wrapper matrices have a positive diagonal entry at a tracked coordinate.
\end{abstract}

\section{Introduction}
\label{sec:intro}

Consider the first-order rewrite rule
\[
  \mathrm{recur}(b,s,\mathrm{succ}(n))
  \;\to\;
  \mathrm{wrap}(s,\mathrm{recur}(b,s,n)).
\]
The step argument $s$ occurs once on the left and twice on the right; we call this duplicated argument the payload. A direct whole-term measure that retains both wrapper arguments must therefore count an additional copy of $s$, while the recursive counter decreases by one. The central question is when the counter decrease dominates the duplicated payload for every rule instance.

\begin{figure}[t]
  \centering
  \resizebox{\linewidth}{!}{%
  \begin{tikzpicture}[
      rec/.style={draw, minimum width=3.1cm, minimum height=0.5cm, inner sep=2pt, font=\footnotesize},
      act/.style={draw, thick, minimum width=3.1cm, minimum height=0.5cm, inner sep=2pt, font=\footnotesize},
      lab/.style={font=\footnotesize}]
    \foreach \i/\X/\c in {0/0/3, 1/4/2, 2/8/1, 3/12/0} {
      \ifnum\i>0
        \foreach \j in {1,...,\i} { \node[rec] at (\X, {0.58*\j}) {$\mathrm{wrap}(s,\;\cdot\;)$}; }
      \fi
      \node[act] at (\X, {0.58*(\i+1)}) {$\mathrm{recur}(b,s,\mathrm{succ}^{\c}\,\mathrm{base})$};
      \node[lab] at (\X, 0) {counter $\c$, wrappers $\i$};
    }
    \foreach \j in {1,2,3} { \node[rec] at (16, {0.58*\j}) {$\mathrm{wrap}(s,\;\cdot\;)$}; }
    \node[act] at (16, {0.58*4}) {$b$};
    \node[lab] at (16, 0) {result, wrappers $3$};
    \foreach \A/\B in {0/4, 4/8, 8/12} {
      \draw[-{Stealth}] ({\A+1.6}, 0.87) -- ({\B-1.6}, 0.87) node[midway, above, font=\scriptsize] {step};
    }
    \draw[-{Stealth}] (13.6, 0.87) -- (14.4, 0.87) node[midway, above, font=\scriptsize] {base};
  \end{tikzpicture}}
  \caption{The trace of the free recursor from counter $3$ (Definition~\ref{def:canonical-trace}). Each application of the duplicating rule removes one successor from the counter and writes one wrapper that keeps the step argument $s$; the base rule replaces the recursor by $b$. The three wrappers are the record of the run, and the counter bounds its length.}
  \label{fig:trace}
\end{figure}

Figure~\ref{fig:trace} shows a run of the two rules from $\mathrm{recur}(b,s,\mathrm{succ}^3\,\mathrm{base})$, the trace of Definition~\ref{def:canonical-trace}. Each step removes one successor from the counter, writes one wrapper that keeps the step argument $s$, and leaves the recursor to act on the smaller counter; when the counter reaches $\mathrm{base}$, the base rule replaces the recursor by $b$. The counter is a natural number, so every run stops after one step for each successor and one base step, and both root and contextual rewriting terminate and are confluent (Theorem~\ref{thm:free-source-normalization}). The wrappers are the record of the run: the result has one wrapper for each step and a copy of $s$ in each. Two rules, one for the step and one for the stop, are enough to record every step of a counted run, and no step rule that writes such a record and continues the recursion can hold the step argument fewer than twice; the companion paper proves this for every step rule of its many-sorted record-emitter syntax whose right side contains a frame and an active site~\cite{rahnama2026operational}. Proof assistants use this system to show that their programs stop: Lean's kernel has no termination checker; Lean derives a recursor of this shape for every inductive type, elaborates every recursive definition into uses of these recursors before the kernel sees it, and the kernel type-checks those uses~\cite{LeanReference}. Theorem~\ref{thm:system-t-iterator} shows that the iterator of G\"odel's System~T~\cite{Godel58} is this two-rule system up to an isomorphism of rewrite relations. Two results in term rewriting concern rules that copy a variable: no standard first-order Knuth-Bendix order orients such a rule~\cite{BaaderNipkow98}, and the disjoint union of two terminating systems can fail to terminate when one of them contains such a rule~\cite{Toyama87}. The question of this paper is which direct measures prove that this system stops.

We answer this question first on the free recursor, the duplicating rule together with the base rule $\mathrm{recur}(b,s,\mathrm{base})\to b$ over free terms, and then transport the mechanism to broader classes. The main characterization concerns a scalar grammar of measure expressions built from two observations, the recursion counter and payload mass, using constants, addition, multiplication, pointwise maximum, and natural scalar multiplication. Every numerical transition needed by the grammar occurs on an actual successor rule instance. Call an expression \emph{payload-independent} when its value at each counter ignores the payload, and \emph{counter-strict} when its value increases strictly with the counter.

\begin{quote}
  \emph{A scalar grammar measure orients every root successor step of the free recursor if and only if it is payload-independent and counter-strict} (Theorem~\ref{thm:orientation-boundary-actual}).
\end{quote}

The free recursor is strongly normalizing and confluent under all contexts (Theorem~\ref{thm:free-source-normalization}), yet every scalar grammar measure that reads the payload fails to orient it, as does every measure of the twelve direct families of Section~\ref{sec:barrier} under their stated hypotheses. One comparison, between the cost of the copied payload and the gain of one counter step, accounts for all of these failures. The two sides of the theorem describe one set of expressions twice, once by the rewrite steps and once by two conditions on the expression alone, and we call this set the Orientation Boundary. Among counter-strict expressions, payload independence is the remaining condition.

The proof mechanism is independent of that grammar. For any natural-valued measure, orientation is equivalent to the inequality
\[
  \mathrm{cost}_M(b,s,n)<\mathrm{gain}_M(b,s,n),
\]
where the left side measures the extra wrapper contribution and the right side measures the change produced by one successor in the counter. If wrapper cost is unbounded in $s$ while counter gain is bounded, orientation fails. In the four-cell classification, the unbounded-cost/bounded-gain cell decides failure by itself; each of the other three cells contains both orienters and non-orienters. This cost/gain view organizes the additive, affine, polynomial, max-plus, vector, matrix, and symbolic-order barriers proved later. For natural-valued measures its global form is one law: a measure that keeps both wrapper arguments at a fixed cost and gains at most a fixed amount per counter step fails to orient the rule (Theorem~\ref{thm:direct-duplication-law}). Additive interpretations satisfy the law, and affine and natural-matrix interpretations with positive wrapper coefficients and finite-diagonal arctic interpretations satisfy instances of it, which is why they fail. A tropical interpretation violates wrapper retention, and a one-dimensional one orients the root rule; no finite-dimensional tropical interpretation orients the rule under both wrapper contexts (Theorem~\ref{thm:tropical-context-barrier}).

Both hypotheses of the failing cell are needed, and each of the twenty premises consumed by the twelve family theorems has a countermodel that satisfies the other premises, violates that one, and orients the rule (Theorem~\ref{thm:premise-necessity}). Seventy-six termination methods from the literature, from Knuth-Bendix orders and matrix interpretations to dependency-pair processors, sized types, and term graph rewriting, receive a termination verdict on the free recursor with typed data, laws, and a machine-checked result (Theorem~\ref{thm:method-classification}).

Orientation and termination are distinct questions for the same rule. We call a termination proof that succeeds by violating a hypothesis of the direct barriers or by using information outside the direct measure classes an \emph{escape}. Dependency pairs project away the duplicated payload and keep only the decreasing counter. A nonlinear interpretation couples counter and payload growth and therefore leaves the transparent-successor class. A fixed-signature path order uses the subterm property and therefore adds structure beyond the direct numerical grammar. Each construction changes the information or structure available to the proof.

\paragraph{Relation to prior work.}
The variable condition prevents standard Knuth-Bendix orders from orienting duplicating rules~\cite{BaaderNipkow98}, while the dependency-pair framework moves termination reasoning from whole rules to recursive calls~\cite{ArtsGiesl00}. Mitterwallner, Middeldorp, and Thiemann~\cite{MitterwallnerMiddeldorpThiemann24,MitterwallnerMiddeldorpThiemannAFP24} formalized undecidability of orientation by linear polynomial interpretations and by KBO with subterm coefficients. Their results quantify over all input systems and ask whether a method class admits a decision procedure. Our main negative results fix a duplicating rule schema and prove non-orientation for every carrier and interpretation satisfying the stated laws. These are different quantifier orders.

Fixed-system orientation results provide a second point of comparison. Dick, Kalmus, and Martin give a complete algorithm for KBO-orientability of a fixed finite rule set~\cite{DickKalmusMartin90}; Podelski and Rybalchenko give complete synthesis for linear ranking functions over linear-arithmetic loops~\cite{PodelskiRybalchenko04}; and Yolcu, Aaronson, and Heule prove dimension-uniform natural-matrix non-orientability for a fixed unary Collatz TRS~\cite{YolcuAaronsonHeule23}. Our contribution is specific to the step-duplicating recursor: the payload-independent/counter-strict characterization, its growth-cell mechanism, the free-term attainment theorem, the associated decision boundary for constructor-local polynomials, and the separation between direct orientation and transformed or structural termination proofs.

Formalization and certification systems such as IsaFoR/CeTA~\cite{IsaFoR}, CoLoR~\cite{CoLoR}, and related developments establish or check positive termination arguments for given systems. The Lean~4 development behind this paper checks impossibility theorems quantified over every schema, every carrier, and every interpretation of the stated shape, and it contains generic infrastructure of the kind those libraries hold: the recursive path order with status and its well-foundedness over every signature, its variants modulo permutative and associative-commutative congruence, a fragment of the higher-order recursive path order on applicative terms, many-sorted persistence, and the soundness of dependency-pair processors (Section~\ref{sec:formalization}). TTT2 and CeTA certify the exported free recursor and the Concrete Witness System (CWS), the fixed seven-constructor, eight-rule system of Section~\ref{sec:cws}, and the barrier theorems predict the MAYBE answers of the Knuth-Bendix strategy of TTT2 on CWS, and of its matrix strategies whenever both wrapper matrices have a positive diagonal entry at a tracked coordinate (Section~\ref{sec:ttt2}).

\paragraph{Contributions.}
The paper makes five contributions.
\begin{enumerate}
  \item \textbf{The Orientation Boundary.} On the free recursor, a scalar grammar measure orients every root successor step if and only if it is payload-independent and counter-strict (Theorem~\ref{thm:orientation-boundary-actual}). The theorem is the equality of two sets of grammar expressions, one defined by the rewrite steps and one by two conditions on the expression, and an attainment theorem realizes every numerical profile of the classification by an actual rewrite step (Theorem~\ref{thm:free-profile-attainment}). The payload-independence half extends to finite-dimensional vector measures under scalar-domination and grammar-agreement hypotheses (Theorem~\ref{thm:vector-grammar-closure}).
  \item \textbf{The barrier mechanism and its tightness.} Wrapper cost against counter gain gives an orientation criterion for every natural-valued measure on every schema, a four-cell growth classification in which one cell decides failure, and one law, wrapper retention with bounded counter gain, that excludes orientation without any shape hypothesis (Theorems~\ref{thm:orientation-criterion}, \ref{thm:four-cells}, and~\ref{thm:direct-duplication-law}). The twelve direct families of Section~\ref{sec:barrier} lie in the failing cell, natural-matrix interpretations whose two wrapper matrices have a positive diagonal entry at a tracked coordinate fail in every finite dimension without a certificate, and every order with the variable condition fails at the rule itself, with or without subterm coefficients. Both hypotheses of the failing cell and every consumed premise of every family theorem have a one-premise countermodel (Theorem~\ref{thm:premise-necessity}).
  \item \textbf{Escapes and seventy-six methods.} Dependency-pair projection, nonlinear coupling, and fixed-signature path orders orient the rule, and every successful direct root-step orienter loses wrapper sensitivity, loses successor transparency, or lies outside the explicit twelve-family universe (Theorem~\ref{thm:escape-trichotomy}). A one-dimensional tropical interpretation orients every root instance, and no finite-dimensional tropical interpretation orients the rule under both wrapper contexts (Theorem~\ref{thm:tropical-context-barrier}). Seventy-six termination methods from the literature receive a termination verdict on the free recursor, each with typed data, laws, a machine-checked result, and a necessity statement (Theorem~\ref{thm:method-classification}).
  \item \textbf{Polynomial decision.} Orientation of the two free-recursor rules by constructor-local natural polynomial interpretations of arbitrary degree has a total decision procedure at unit successor slope, with a failing assignment on rejection (Theorem~\ref{thm:tdec1}). At higher slopes the undecided remainder of the procedure is a signed positivity problem that contains the zero problem for integer polynomials in two variables over $\mathbb N^2$, whose decidability is open~\cite{Poonen08}, and six named residual subclasses are decided (Theorems~\ref{thm:residual-bivariate} and~\ref{thm:residual-strata}).
  \item \textbf{Termination, complexity, and transport.} The free recursor is strongly normalizing and confluent under all contexts (Theorem~\ref{thm:free-source-normalization}). Its contextual derivation length is at most $3^{|t|}-2$, and a family of terms of size $5d+4$ has derivations of length at least $2^{d}$ (Theorem~\ref{thm:free-exponential}). Its escapes are calibrated at $\omega$ and $\varphi_4(0)$ (Theorem~\ref{thm:free-ordinal-height}). Its root derivation height is the least natural-valued root ranking, and the counter and payload observations do not determine it (Theorem~\ref{thm:free-profile-attainment}). Four results hold over every signature that contains the four schema symbols (Theorem~\ref{thm:signature-transport}): the failing-cell theorem, the bound $M(s)+c_w<\mathrm{gain}_M(b,s,n)$ for orienters that retain both wrapper arguments at a fixed cost $c_w$, contextual termination for interpretations into a well-founded order that orient both root rules and are strictly monotone in every argument, and the unit-slope decision procedure. The orientation criterion holds at every duplication arity, and the barrier at every positive one (Theorem~\ref{thm:arity-coupling}). For every sort assignment that types its rules, sorted termination of the free recursor is equivalent to unsorted strong normalization (Theorem~\ref{thm:free-recursor-sorted-persistence}). In System~T the iterator fragment is the free recursor up to an isomorphism of rewrite relations, and the recursor rule copies both the step argument and the counter (Theorems~\ref{thm:system-t-iterator} and~\ref{thm:system-t-recursor}). The CWS case study supplies a fixed system with relation-specific termination, confluence, normalization, complexity, and nonconfluence results (Section~\ref{sec:certification}).
\end{enumerate}
Every numbered result is machine-checked in Lean~4 (Section~\ref{sec:formalization}). The boundary, the growth cells, the premise slots, and the method rows are formal objects of the development, which also contains every family theorem with its violating triple, the control examples, and the local-join data; the claim-to-code table of the public artifact indexes them.

\begin{table}[ht]
  \centering
  \small
  \caption{Scope of the main result groups.}\label{tab:scope-matrix}
  \begin{tabularx}{\textwidth}{@{}P{0.18\linewidth}P{0.23\linewidth}P{0.25\linewidth}X@{}}
    \toprule
    Result group & Relation / carrier & Quantifier domain & Main conclusion and hypotheses \\
    \midrule
    Scalar Orientation Boundary & Actual free-recursor successor root steps & Every reflected scalar grammar expression & Orientation if and only if payload-independent and counter-strict; on counter-strict expressions, if and only if payload-independent. \\
    Vector closure & Finite-dimensional vector measures & Every finite dimension with a grammar-expressible scalarization & Orientation forces payload independence of the scalarization under domination and grammar agreement. \\
    Direct duplication law & Every step-duplicating schema & Every natural-valued measure that retains both wrapper arguments with bounded counter gain & Uniform orientation fails. \\
    Premise necessity & Twelve direct families, free schema & Twenty consumed premises & Each premise has a one-premise countermodel. \\
    Escape trichotomy & Direct root-step orienters & Explicit finite direct universe, with a separate projection extension & Success requires loss of wrapper sensitivity, loss of successor transparency, or exit from that universe. \\
    Method classification & Free recursor & Seventy-six named termination methods & Each row has typed data, laws, and a proved termination verdict; fourteen rows use a removable barrier premise, seven reject by a law-free obstruction, forty-two change termination verdict with the escape datum, thirteen have a termination verdict fixed by their laws. \\
    Polynomial decision & Fixed free two-rule root system & Constructor-local natural polynomials of arbitrary degree & Total decision at unit successor slope; higher slopes reduce to a signed residual with six decidable subclasses. \\
    Arbitrary semantic erasure & Abstract RDRS root step & Every semantic measure on a step carrying a static payload erasure & Every orienting measure is counter-dominated; decisive payload-sensitive orientation is impossible. \\
    Finite semantic classifier & Closed normalized certificates and reflected direct syntax & Declared finite inductive syntax & Total classification of that certificate or syntax universe. \\
    CWS case study & Root, guarded, and contextual CWS relations & Fixed CWS system & Relation-specific termination, normalization, confluence, complexity, and nonconfluence results. \\
    External validation & Pinned CWS and free-recursor TPDB/CPF problems & Fixed exported TRS objects & TTT2/CeTA certification of the CWS context-termination instance and of the two-rule free recursor. \\
    \bottomrule
  \end{tabularx}
\end{table}

\paragraph{Organization.}
Section~\ref{sec:prelim} fixes the schema, the trace of the recursor, and the running systems. The cost/gain criterion, the growth cells, and the direct duplication law come first in Section~\ref{sec:criterion}, followed by the Orientation Boundary, whose proof combines a structural induction over the grammar with the attainment of every numerical profile by a term. Section~\ref{sec:barrier} applies the criterion to the direct families, proves that their consumed premises are necessary, constructs the escapes, proves that the free recursor terminates, and classifies seventy-six termination methods by the mechanism that decides each termination verdict; Appendix~\ref{sec:methods} lists the methods with their citations. From the criterion, Section~\ref{sec:decision} derives a decision procedure for the orientation of the two free-recursor rules by constructor-local natural polynomial interpretations at unit successor slope, and shows that at higher slopes its undecided remainder contains the zero problem for two-variable integer polynomials over $\mathbb N^2$. The later sections use these results in four directions. Section~\ref{sec:schema-diagnostic} treats arbitrary semantic measures: when a map removes the payload from both sides of a step and keeps the base and the counter, composing an orienting measure with that map gives a payload-independent measure that still orients. Section~\ref{sec:costs} measures the derivation lengths of the free recursor and the ordinals its termination proofs use. Transport to other signatures, arities, and System~T is the subject of Section~\ref{sec:transport}, and Sections~\ref{sec:certification} and~\ref{sec:ttt2} study CWS and its external certification. Section~\ref{sec:formalization} describes the Lean development and the public artifact.

\section{Background}
\label{sec:background}

We use the standard notions of abstract reduction and term rewriting~\cite{BaaderNipkow98,Terese03}. A term is in \emph{normal form} when every rule is inapplicable. Strong normalization (SN) means that every reduction sequence is finite. Confluence means that for any $t \Rightarrow^{\ast} u$ and $t \Rightarrow^{\ast} v$ there exists $w$ with $u \Rightarrow^{\ast} w$ and $v \Rightarrow^{\ast} w$. \emph{Local confluence} restricts this condition to single-step forks. \emph{Newman's Lemma} states that SN and local confluence together imply confluence~\cite{Newman42}; under SN and confluence every term has a unique normal form.

\section{Preliminaries: schema, concrete calculus, and \texorpdfstring{Rec$\Delta$}{RecDelta}-core}
\label{sec:prelim}

\subsection{The step-duplicating schema}\label{sec:schema}

\begin{definition}[Step-duplicating schema]\label{def:schema}
  A \emph{step-duplicating schema} consists of a term carrier $T$ with designated operations
  \[
    \mathrm{base}\in T,\qquad \mathrm{succ}:T\to T,\qquad
    \mathrm{wrap}:T\times T\to T,\qquad \mathrm{recur}:T^3\to T,
  \]
  and a step relation containing the duplicating rule
  \[
    \mathrm{recur}(b,s,\mathrm{succ}(n)) \to \mathrm{wrap}(s,\mathrm{recur}(b,s,n)).
  \]
\end{definition}

We call Definition~\ref{def:schema}, with its position and occurrence
data, the \emph{Right-Duplicating Recursor Schema} (RDRS). The definition
fixes which argument is duplicated and which is the counter; the term
carrier and any additional rewrite rules stay arbitrary.

\begin{definition}[Direct compositional measures]\label{def:compositional}
  Two compositional axiom systems, both mapping terms to $\mathbb{N}$:
  \begin{enumerate}
    \item \textbf{Tier~1 (additive).}
      A fixed weight per constructor; the measure of a compound term is the constructor's weight plus the sum of subterm measures. Hypothesis: \texttt{wrap} contributes at least~1.

    \item \textbf{Tier~2 (abstract, with wrapper-subterm axioms).}
      Each constructor has an arbitrary combining function on subterm measures, subject to:
      \[
        c_{\mathrm{wrap}}(x,y) > x
        \qquad\text{and}\qquad
        c_{\mathrm{wrap}}(x,y) > y.
      \]
      Tier~1 permits arbitrary growth behavior for $\mathrm{succ}$. The Tier~2 impossibility also requires $\mathrm{succ}$-transparency at $\mathrm{base}$: $c_{\mathrm{succ}}(c_{\mathrm{base}}) = c_{\mathrm{base}}$.
  \end{enumerate}
  A measure \emph{orients} a rule instance if the measured value strictly decreases across that instance.
\end{definition}

\noindent Theorem~\ref{thm:affine-barrier} isolates an affine/linear constructor-local fragment:
\[
  M(\mathrm{succ}(t)) = \alpha_s + \beta_s\, M(t),\qquad
  M(\mathrm{wrap}(x,y)) = \alpha_w + \beta_w\, M(x) + \gamma_w\, M(y),
\]
\[
  M(\mathrm{recur}(b,s,n)) = \alpha_r + \beta_r\, M(b) +
  \gamma_r\, M(s) + \delta_r\, M(n),
\]
with every cross-term coefficient equal to zero. The constructor equations supply the bound used in its barrier proof.

\noindent The core schema and barrier theorems below concern one self-recursive duplicating rule. The same mechanism extends to explicit two-rule and finite-cycle mutual schemas: one complete cycle yields the same additive and affine obstruction, together with transparent-compositional and scalar-projection versions of it. Other mutual systems require their own cycle or payload-flow hypotheses.

\begin{remark}[Classic duplicating TRS as schema instance]\label{rem:classic-instance}
  The schema also captures standard parameterized duplicating rules from first-order rewriting. For example, the rule
  \[
    f(x,\,s(y)) \;\to\; g(x,\,f(x,y))
  \]
  is a direct instance: take $\mathrm{base}=b$ as a dummy parameter, let $\mathrm{succ}=s$ act on the second argument, set $\mathrm{wrap}(a,r)=g(a,r)$, and define $\mathrm{recur}(b,a,n)=f(a,n)$. Then the schema step
  \[
    \mathrm{recur}(b,a,\mathrm{succ}\,n)\;\to\;\mathrm{wrap}(a,\mathrm{recur}(b,a,n))
  \]
  becomes the displayed rule. Every barrier proved below therefore applies to this duplicating shape under the hypotheses stated in the corresponding theorem, and in the affine case the wrapper coefficients supply the needed growth. The obstruction follows from the self-recursive duplication of the step argument alone.
\end{remark}

\subsection{The trace, its coordinates, and the wrapper stack}\label{sec:canonical-trace}

Add the base rule $\mathrm{recur}(b,s,\mathrm{base})\to b$ to a step-duplicating schema. The resulting \emph{base-duplicating system} is the abstract version of the first-order pattern $F(x,y,Z)\to x$, $F(x,y,S(n))\to G(y,F(x,y,n))$. The next definition fixes one sequence of terms from $\mathrm{recur}(b,s,\mathrm{succ}^k\,\mathrm{base})$ to $\mathrm{wrap}^k\,s\,b$, and the definitions after it read three coordinates along that sequence.

\begin{definition}[Trace]\label{def:canonical-trace}
  Fix a base-duplicating system and $b, s \in T$. For $k \ge 0$, the \emph{trace state} at step $i \le k$ starting from $\mathrm{recur}(b, s, \mathrm{succ}^k\,\mathrm{base})$ is
  \[
    t_i \;:=\; \mathrm{wrap}^i\,s\,(\mathrm{recur}(b, s, \mathrm{succ}^{k-i}\,\mathrm{base})),
  \]
  followed by the terminal state $t_{k+1}:= \mathrm{wrap}^k\,s\,b$ reached by one application of the base rule. The \emph{trace} is the sequence $t_0,\dots,t_{k+1}$, and the \emph{trace step} at stage $i < k$ is the duplicating rule firing at the root of the innermost $\mathrm{recur}$ site.
\end{definition}

\begin{definition}[Trace coordinates]\label{def:trace-coords}
  For every live stage $0 \le i \le k$, the counter, payload-multiplicity, and wrapper-multiplicity coordinates along the trace are
  \[
    \mathrm{ctr}(t_i):= k - i,
    \qquad
    \mathrm{pay}(t_i):= i + 1,
    \qquad
    \mathrm{wraps}(t_i):= i.
  \]
\end{definition}

\begin{proposition}[Per-step control/payload exchange]\label{prop:per-step-exchange}
  For every $i < k$,
  \[
    \mathrm{ctr}(t_i) = \mathrm{ctr}(t_{i+1}) + 1
    \qquad\text{and}\qquad
    \mathrm{pay}(t_{i+1}) = \mathrm{pay}(t_i) + 1.
  \]
  Each trace step consumes one unit of counter structure and creates one additional payload slot.
\end{proposition}

\begin{proposition}[Offset conservation]\label{prop:offset-conservation}
  For every live stage $0 \le i \le k$,
  \[
    \mathrm{pay}(t_i) = \mathrm{wraps}(t_i) + 1.
  \]
  The payload-multiplicity coordinate and the wrapper-multiplicity coordinate differ by one at every live stage. On the formal trace the same law holds for counts: the number of payload occurrences equals the number of emitted wrapper constructors plus one.
\end{proposition}

\begin{definition}[Wrapper cell and wrapper stack]\label{def:wrapper-cell}
  Fix a size function $|\cdot|:T\to\mathbb N$. The \emph{wrapper-cell weight} is $w:=|\mathrm{wrap}|+|s|$, the structural size of one wrapper-payload pair. The \emph{wrapper stack} at step $i$ is the length-$i$ list of identical payload copies held in the accumulating wrapper frames. Its length satisfies $|\mathrm{stack}_i|+1=\mathrm{pay}(t_i)$.
\end{definition}

\begin{proposition}[Counter is the permutation-invariant retained coordinate]\label{prop:counter-retained}
  At step $i < k$ the three coordinates $(\mathrm{ctr}(t_i), \mathrm{pay}(t_i), \mathrm{wraps}(t_i))$ satisfy $\mathrm{ctr}(t_{i+1}) < \mathrm{ctr}(t_i)$ (strict descent), $\mathrm{pay}(t_i) < \mathrm{pay}(t_{i+1})$ (strict ascent), and $\mathrm{wraps}(t_i) < \mathrm{wraps}(t_{i+1})$ (strict ascent). Under any permutation of the $i+1$ payload positions at stage $i$, the constant payload tuple is fixed pointwise, while the counter $\mathrm{succ}^{k-i}\,\mathrm{base}$ is fixed because it is payload-free. So $\mathrm{ctr}$ is the one coordinate that is both permutation-invariant and strictly descending.
\end{proposition}

\noindent Theorem~\ref{thm:grammar-closure-barrier} reads the trace step in mass coordinates, where one duplicating firing takes counter $c+1$ to $c$ and adds the payload mass of the copy.

\subsection{The Concrete Witness System}\label{sec:cws}

The Concrete Witness System is a first-order ground term calculus with 7 constructors and 8 rules.

\paragraph{Kernel signature}
\[
  t::= \mathrm{void}
  \mid \delta\,t
  \mid \mathrm{integrate}\,t
  \mid \mathrm{merge}\,t\,t
  \mid \mathrm{app}\,t\,t
  \mid \mathrm{rec}\Delta\,t\,t\,t
  \mid \mathrm{eqW}\,t\,t.
\]

\begin{table}[ht]
  \centering
  \caption{The eight root rules of the concrete calculus. The recursor successor rule ($*$) duplicates the step argument~$s$. In the unguarded relation the unequal-case equality rule is available without the guarded side condition.}
  \label{tab:cws-rules}
  \begin{tabular}{@{}llclc@{}}
    \toprule
    Rule & Head & Arity & Shape & Additional copies \\
    \midrule
    merge-L & merge & 2 & $\mathrm{merge}\;\mathrm{void}\;t \to t$ & 0 \\
    merge-R & merge & 2 & $\mathrm{merge}\;t\;\mathrm{void} \to t$ & 0 \\
    merge-idem & merge & 2 & $\mathrm{merge}\;t\;t \to t$ & 0 \\
    rec-zero & rec$\Delta$ & 3 & $\mathrm{rec}\Delta\,b\,s\,\mathrm{void} \to b$ & 0 \\
    rec-succ & rec$\Delta$ & 3 & $\mathrm{rec}\Delta\,b\,s\,(\delta\,n) \to \mathrm{app}\,s\,(\mathrm{rec}\Delta\,b\,s\,n)$ & $1^{*}$ \\
    integrate & integrate & 1 & $\mathrm{integrate}\,(\delta\,t) \to \mathrm{void}$ & 0 \\
    eq-refl & eqW & 2 & $\mathrm{eqW}\;a\;a \to \mathrm{void}$ & 0 \\
    eq-diff & eqW & 2 & $\mathrm{eqW}\;a\;b \to \mathrm{integrate}\,(\mathrm{merge}\,a\,b)$ & 0 \\
    \bottomrule
  \end{tabular}
\end{table}

\noindent In the unguarded root relation, the two equality rules overlap at $\mathrm{eqW}(a,a)$. The guarded relation adds rule-specific side conditions: $\delta$-flag guards on merge-with-void and rec-zero support termination; $\delta\text{-flag}(t)=0$ and $\kappa^M(t)=\varnothing$ control idempotent merge collapse; and $\kappa^M(a)=\varnothing$ on the reflexive equality rule together with $a\neq b$ on the unequal-case rule separate the equality peak.

\paragraph{Relations}
\medskip
\begin{center}
  \captionof{table}{Certified properties of the primary concrete reduction relations.}
  \label{tab:relations-overview}
  \small
  \begin{tabularx}{\textwidth}{@{}l l c Y Y@{}}
    \toprule
    Relation & Scope & SN & Confluence & Normalizer \\
    \midrule
    $R_{\mathrm{root}}$ & Full, root & Terminating & Nonconfluent & Two terminating rule-priority normalizers \\
    $R_{\mathrm{safe}}$ & Guarded, root & Strongly normalizing & Confluent & Unique normal forms \\
    $R_{\mathrm{safe}}^{\mathrm{ctx}}$ & Partial context closure & Strongly normalizing & Confluent & Contextual normalization \\
    \bottomrule
  \end{tabularx}
\end{center}

\noindent The guards of \texttt{SafeStep} serve confluence: the unguarded root relation \texttt{Step} also terminates (Proposition~\ref{prop:poly-full-step}). The functions \texttt{deltaFlag} and $\kappa^M$ appear both in the guarded rules and in the triple-lexicographic decrease argument of the computable termination certificate. The relation \texttt{Step} fails local confluence at the concrete term $\mathrm{eqW}\;\mathrm{void}\;\mathrm{void}$:
\[
  \mathrm{eqW}\;\mathrm{void}\;\mathrm{void}\to \mathrm{void}
  \qquad\text{and}\qquad
  \mathrm{eqW}\;\mathrm{void}\;\mathrm{void}\to \mathrm{integrate}(\mathrm{merge}\;\mathrm{void}\;\mathrm{void}).
\]
Both targets are root normal forms of \texttt{Step} and form a nonjoinable fork. The guarded relation \texttt{SafeStep} removes this overlap, and its root abstract reduction system has unique normal forms.

\noindent The constructor $\mathrm{app}$ is irreducible. When the successor recursor rule fires, the duplicated $s$ in $\mathrm{app}(s,\mathrm{rec}\Delta(b,s,n))$ remains outside the recursion-counter position because every left-hand side has a root distinct from $\mathrm{app}$. The duplicated payload therefore remains outside any additional root recursive control position.

\subsection{\texorpdfstring{Rec$\Delta$}{RecDelta}-core}\label{par:rec-core}
The impossibility and DP escape depend only on four constructors and two rules:
\[
  \begin{array}{rcl}
    t &::=& \mathrm{void} \mid \delta\,t \mid \mathrm{app}\,t\,t \mid \mathrm{rec}\Delta\,t\,t\,t \\[4pt]
    \mathrm{rec}\Delta\,b\,s\,\mathrm{void} &\to& b \\
    \mathrm{rec}\Delta\,b\,s\,(\delta\,n) &\to& \mathrm{app}\,s\,(\mathrm{rec}\Delta\,b\,s\,n)
  \end{array}
\]
The concrete calculus extends Rec$\Delta$-core with \texttt{merge}, \texttt{integrate}, and \texttt{eqW} for the confluence rules. The barrier holds within Rec$\Delta$-core.

\section{The orientation criterion}
\label{sec:criterion}

The duplicating instance at $(b,s,n)$ compares the value of $\mathrm{recur}(b,s,\mathrm{succ}(n))$ with the value of $\mathrm{wrap}(s,\mathrm{recur}(b,s,n))$. Both sides are measured against the term $\mathrm{recur}(b,s,n)$: the left side is that term with one more successor on its counter, and the right side contains it as the second wrapper argument. Subtracting its value from both sides leaves two integers, and orientation becomes a comparison between them. Every result of this section before \S\ref{sec:exact-orientation-boundary} holds for an arbitrary step-duplicating schema (Definition~\ref{def:schema}) and an arbitrary function $M:T\to\mathbb{N}$, except where its statement names the free schema, and each statement names every constructor equation, monotonicity law, or growth premise it uses.

\begin{definition}[Wrapper cost and counter gain]\label{def:cost-gain}
  For $M:T\to\mathbb{N}$ and $b,s,n\in T$, the \emph{wrapper cost} and the \emph{counter gain} at $(b,s,n)$ are the integers
  \begin{align*}
    \mathrm{cost}_M(b,s,n)&=M(\mathrm{wrap}(s,\mathrm{recur}(b,s,n)))-M(\mathrm{recur}(b,s,n)),\\
    \mathrm{gain}_M(b,s,n)&=M(\mathrm{recur}(b,s,\mathrm{succ}(n)))-M(\mathrm{recur}(b,s,n)).
  \end{align*}
\end{definition}

\begin{theorem}[Orientation criterion]\label{thm:orientation-criterion}
  For every step-duplicating schema, every $M:T\to\mathbb{N}$, and every triple $(b,s,n)$, the value of $M$ strictly decreases across the duplicating instance at $(b,s,n)$ if and only if $\mathrm{cost}_M(b,s,n)<\mathrm{gain}_M(b,s,n)$. Hence $M$ orients the duplicating rule if and only if the wrapper cost lies below the counter gain at every triple.
\end{theorem}

\noindent Both comparisons differ by the same integer $M(\mathrm{recur}(b,s,n))$, and integer subtraction keeps the equivalence free of truncation. The criterion becomes informative once a measure class constrains the two functions. The first constraint is retention: for a measure that keeps both wrapper arguments, orientation forces the counter gain above the value of the duplicated copy of $s$.

\begin{theorem}[Coupling inequality]\label{thm:coupling-inequality}
  Let $M:T\to\mathbb{N}$ and $c_w\in\mathbb{N}$ satisfy $c_w+M(x)+M(y)\le M(\mathrm{wrap}(x,y))$ for all $x,y\in T$. If $M$ orients the duplicating rule, then
  \[
    M(s)+c_w<\mathrm{gain}_M(b,s,n)\qquad\text{for all }b,s,n\in T.
  \]
  Hence, if the counter gain is bounded by a constant $g$ at every triple and $M$ takes values above every bound, then $M$ fails to orient the duplicating rule.
\end{theorem}

\noindent The inequality is attained. Under the interpretation $w(\mathrm{succ}(n))=n+1$, $w(\mathrm{wrap}(s,y))=s+y+1$, and $w(\mathrm{recur}(b,s,n))=(n+1)(s+b+2)$ at base value zero, the counter gain equals $s+2$ at every counter and the retention cost is one, so the strict inequality of Theorem~\ref{thm:coupling-inequality} holds with a margin of one unit. The unboundedness premise of the second sentence is dispensable: orientation itself supplies the unbounded values.

\begin{theorem}[Direct duplication law]\label{thm:direct-duplication-law}
  Let $M:T\to\mathbb N$ on a step-duplicating schema and let $c_w,g\in\mathbb N$ satisfy
  \begin{gather*}
    c_w+M(x)+M(y)\le M(\mathrm{wrap}(x,y))\qquad\text{for all }x,y,\\
    M(\mathrm{recur}(b,s,\mathrm{succ}(n)))\le g+M(\mathrm{recur}(b,s,n))\qquad\text{for all }b,s,n.
  \end{gather*}
  Then $M$ fails uniform strict orientation of the duplicating step, and so does every relation on $\mathbb N$ whose strict comparison implies strict decrease of $M$, on every rewrite relation that contains the step. No linearity, compositionality, matrix or polynomial shape, or range hypothesis is used.
\end{theorem}
\begin{proof}
  Suppose $M$ orients every instance. At the base triple the target is at least $c_w$, so the seed $u=\mathrm{recur}(\mathrm{base},\mathrm{base},\mathrm{succ}(\mathrm{base}))$ has $M(u)\ge1$. Wrapping the seed around itself $g$ times gives a term $t$ with $M(t)\ge g+1$ by the retention clause. Orientation at $(\mathrm{base},t,\mathrm{base})$ and retention give $M(t)<M(\mathrm{recur}(\mathrm{base},t,\mathrm{succ}(\mathrm{base})))-M(\mathrm{recur}(\mathrm{base},t,\mathrm{base}))\le g$, a contradiction.
\end{proof}

\noindent Every additive interpretation satisfies the law, and the affine, natural-matrix, finite-diagonal arctic, and ordered-carrier families satisfy instances of it; this is the mechanism behind their barriers in Section~\ref{sec:barrier}. The tropical family violates wrapper retention, which is why a min-plus interpretation orients the root rule (Theorem~\ref{thm:tropical-context-barrier}). Both clauses are used: the counter projection satisfies the gain clause, violates retention, and orients; the coupled interpretation displayed above satisfies retention, violates the gain clause, and orients.

\smallskip\noindent The second constraint concerns growth in the duplicated argument. Fix a base $b$ and a counter $n$ and let the payload $s$ range over $T$. The wrapper cost and the counter gain become functions of $s$, and each is bounded above or unbounded.

\begin{definition}[Growth cells]\label{def:growth-cells}
  At a fixed pair $(b,n)$, a function $M:T\to\mathbb{N}$ has \emph{unbounded wrapper cost} if for every $K$ some $s$ satisfies $M(\mathrm{recur}(b,s,n))+K<M(\mathrm{wrap}(s,\mathrm{recur}(b,s,n)))$, and \emph{bounded wrapper cost} if some $K$ satisfies $M(\mathrm{wrap}(s,\mathrm{recur}(b,s,n)))\le M(\mathrm{recur}(b,s,n))+K$ for every $s$. It has \emph{unbounded} or \emph{bounded counter gain} by the same two clauses with $\mathrm{recur}(b,s,\mathrm{succ}(n))$ in place of the wrapped term. The four combinations are the \emph{growth cells} at $(b,n)$, and the cell with unbounded wrapper cost and bounded counter gain is the \emph{barrier cell}.
\end{definition}

\begin{theorem}[Four-cell classification]\label{thm:four-cells}
  Let $M:T\to\mathbb{N}$ on a step-duplicating schema and fix a pair $(b,n)$.
  \begin{enumerate}
    \item $M$ lies in one of the four growth cells at $(b,n)$.
    \item If $M$ lies in the barrier cell at $(b,n)$, then some payload $s$ satisfies $M(\mathrm{recur}(b,s,\mathrm{succ}(n)))<M(\mathrm{wrap}(s,\mathrm{recur}(b,s,n)))$. In particular $M$ fails to orient the duplicating rule, and it fails even the nonstrict comparison at that pair.
    \item On the free schema, each of the three remaining cells contains a measure that orients the duplicating rule and a measure that fails to orient it.
  \end{enumerate}
\end{theorem}

\noindent For (ii), let $K$ bound the counter gain and choose $s$ whose wrapper cost exceeds $K$; Theorem~\ref{thm:orientation-criterion} then reverses the comparison. For (iii), Table~\ref{tab:cell-witnesses} lists six interpretations of the free schema, all with base value zero and successor $n\mapsto n+1$. The table also shows that the successful measures occupy three cells and share each with failing ones: an orienter can have bounded wrapper cost together with unbounded counter gain. The barrier cell is the only cell whose membership decides the outcome.

\begin{table}[ht]
  \centering\small
  \caption{Every growth cell outside the barrier cell holds an orienter and a non-orienter of the duplicating rule on the free schema.}
  \label{tab:cell-witnesses}
  \begin{tabularx}{\textwidth}{@{}P{3.2cm}YY@{}}
    \toprule
    Cell at $(\mathrm{zero},\mathrm{zero})$ & Orients & Fails \\
    \midrule
    Bounded cost, bounded gain & Counter projection $\pi$ of Theorem~\ref{thm:dp-escape} & Constant zero \\
    Unbounded cost, unbounded gain & $W(s,y)=s+y+1$, $R(b,s,n)=(n+1)(s+b+2)$ & $W(s,y)=2y+s+1$ with the same $R$ \\
    Bounded cost, unbounded gain & $W(s,y)=y+1$, $R(b,s,n)=(n+1)(s+b+2)$ & $W(s,y)=y+1$, $R(b,s,n)=b+\lfloor n/2\rfloor\,s$, at the pair $(\mathrm{zero},\mathrm{succ}(\mathrm{zero}))$ \\
    \bottomrule
  \end{tabularx}
\end{table}

\begin{proposition}[Both hypotheses of the barrier cell are needed]\label{prop:cell-hypotheses-necessary}
  On the free schema, the counter projection has bounded counter gain at $(\mathrm{zero},\mathrm{zero})$, fails unbounded wrapper cost there, and orients the duplicating rule; the coupled interpretation $W(s,y)=s+y+1$, $R(b,s,n)=(n+1)(s+b+2)$ has unbounded wrapper cost, fails bounded counter gain, and orients. Theorem~\ref{thm:four-cells}(ii) is therefore false with either hypothesis deleted.
\end{proposition}

\begin{corollary}[The direct families lie in the barrier cell]\label{cor:families-barrier-cell}
  Every measure of the following families lies in the barrier cell at the base pair $(\mathrm{base},\mathrm{base})$, or fails the duplicating rule at an explicit triple: additive; transparent compositional; affine, restricted quadratic, and max-plus with their wrapper conditions; bounded cross-term quadratic, bounded multilinear, and generalized bounded polynomial under their dominance conditions; and the componentwise, lexicographic, balanced mixed-coordinate, and weighted scalar-projection vector families through their tracked scalar. Each barrier theorem of Section~\ref{sec:barrier} for these twelve families is therefore an instance of Theorem~\ref{thm:four-cells}(ii).
\end{corollary}

\noindent The class lemmas behind the corollary are proved from the family equations and the self-nested wrapper chain; none of them calls the barrier theorem of its own family. Affine measures are identically zero or lie in the barrier cell; max-plus measures have counter gain zero once the payload branch dominates, against a wrapper cost of at least one; transparent compositional measures have counter gain zero and positive wrapper cost at every payload; the three dominance classes lie in the barrier cell when their payload coefficient does not grow from $\mathrm{base}$ to $\mathrm{succ}(\mathrm{base})$ and fail at every payload past a computed bound otherwise. For a scalar grammar measure on the $(\mathrm{counter},\mathrm{payload})$ carrier, orientation holds if and only if the measure avoids the barrier cell at every pair and responds strictly to the counter, which is the growth-cell form of Theorem~\ref{thm:grammar-closure-literal}.

\subsection{Orientation Boundary on the free recursor}\label{sec:exact-orientation-boundary}

The growth-cell analysis becomes a classification on the free recursor once the measure is restricted to the reflected scalar grammar and the numerical profiles are connected back to terms.

\begin{theorem}[Grammar-closure direct-measure barrier]\label{thm:grammar-closure-barrier}
  The scalar negative side closes under the full polynomial-plus-max algebra by a single structural induction.

  Work on the $(\mathrm{counter},\mathrm{payload})$ carrier, where the payload coordinate is payload mass and one duplicating firing takes $(c+1,p)$ to $(c,p+L)$ with $L$ the payload size of the duplicated copy, matching the schema step of \S\ref{sec:canonical-trace} in mass coordinates and the freeze-and-pump construction of Theorem~\ref{thm:schema-barrier}, which fixes the base and the counter and increases the payload without bound. Let the reflected grammar be generated by the counter coordinate, the payload coordinate, constants, sums, products, pointwise maxima, and natural scalar multiples, with the evident denotation into measures on that carrier. Say a measure has \emph{unbounded payload dependence} when, at a fixed counter, its value exceeds every bound as the payload mass grows.

  Then every generated measure with unbounded payload dependence fails to orient the duplicating step uniformly: a fixed left-hand-side value fails to dominate the unbounded family of right-hand-side values produced by the pump. The conclusion follows from the unbounded payload section alone; every grammar expression is payload-monotone. A syntactic test detects effective payload dependence on a sound subclass, but the theorem itself is semantic: whenever the payload section is unbounded, uniform orientation fails. The payload-independent counter projection orients the step, so the same analysis identifies both sides of the boundary.

  This single induction covers the additive, affine, restricted-quadratic, bounded cross-term, generalized polynomial, and max-plus scalar families instead of requiring a separate closure proof for each one. Theorem~\ref{thm:vector-grammar-closure} gives the corresponding finite-dimensional scalarization result.
\end{theorem}

\begin{theorem}[Literal closure and counter-strict biconditional]\label{thm:grammar-closure-literal}
  The barrier of Theorem~\ref{thm:grammar-closure-barrier} admits a hypothesis-free closure over the whole grammar. Call a measure on the $(\mathrm{counter}, \mathrm{payload})$ carrier \emph{payload-independent} when, at every fixed counter, its value is independent of the payload coordinate. Call a measure \emph{counter-strict} when $m(c,p)<m(c+1,p)$ for every $c,p$.

  The grammar (counter, payload, constants, sums, products, pointwise maxima, natural scalar multiples) omits every operation that caps a value from above. A structural induction over its seven constructors, with payload-monotonicity supplying the product and scalar cases, therefore yields an exhaustive dichotomy: at every fixed counter, a generated measure's payload section is either constant or unbounded.

  A grammar measure therefore orients the duplicating step only if it is payload-independent. This implication uses neither an unboundedness hypothesis nor an effective-payload hypothesis: the payload pump refutes orientation at every counter where the section is unbounded. Orientation also forces counter strictness by instantiating the duplicated-payload size at one. These two necessities combine with a converse on the carrier: a grammar expression $e$ orients the duplicating step if and only if $e$ is payload-independent and counter-strict. In the notation of the Lean development,
  \[
    \mathrm{OrientsDupStep}(e.\mathrm{eval})
    \quad\Longleftrightarrow\quad
    \mathrm{PayloadBlind}(e.\mathrm{eval})\ \land\
    \mathrm{CounterStrict}(e.\mathrm{eval}).
  \]
  For the reverse direction, payload independence rewrites $m(c,p+L)$ to $m(c,p)$, and counter strictness supplies the required comparison. Every counter-strict grammar expression therefore orients the step if and only if it is payload-independent. The counter projection supplies an inhabited case. The constant-zero expression is payload-independent, has equal values at adjacent counters, and fails orientation; this shows that the counter-strictness premise is needed.

  A form on the carrier drops the grammar: with no syntax premise, any payload-monotone measure with an unbounded payload section at some counter fails to orient the step, and the grammar statement is one instance. The reflected syntax also gives a computable sufficient test for payload dependence. That test is sound and incomplete: the product of the payload and counter coordinates depends on payload at every positive counter, yet a syntactic zero-erasure test can miss it because the counter factor vanishes at counter zero.
\end{theorem}

\begin{theorem}[Attainment on terms and the least root ranking]
  \label{thm:free-profile-attainment}
  On the free syntax of Theorem~\ref{thm:free-source-normalization}, put
  \[
    \begin{aligned}
      c(\mathrm{recur}(b,s,n))&=\operatorname{sp}(n)+c(b),&
      c(\mathrm{wrap}(s,t))&=c(t),\\
      p(\mathrm{recur}(b,s,n))&=p(b),&
      p(\mathrm{wrap}(s,t))&=p(s)+p(t)+1,
    \end{aligned}
  \]
  and zero at other heads, with $\operatorname{sp}$ the initial successor
  prefix length. Every transition $(c+1,p)\to(c,p+L)$, $L\geq1$, is
  the observed profile of an actual successor rule instance. For every
  $m:\mathbb N^2\to\mathbb N$, orientation of all those term instances
  by $m\circ(c,p)$ is equivalent to the numeric orientation condition.
  Hence the scalar grammar classification applies to those
  actual steps. The executable root height $h$ is the pointwise least
  natural-valued ranking of both root rules, and it cannot factor through
  $(c,p)$.
\end{theorem}
\begin{proof}
  For any $c,p,L$, choose a base argument with $p$ wrappers, a generator
  with $L-1$ wrappers, all with zero left arguments, and counter
  $\mathrm{succ}^c(\mathrm{base})$. These terms attain the four required
  values. The converse follows from the observation equations, so
  Theorem~\ref{thm:grammar-closure-literal} applies without a missing
  attainment premise. A natural ranking decreases by at least one along
  every root path; hence it bounds the length $h$. The height itself drops
  by one. The terms $\mathrm{base}$ and
  $\mathrm{recur}(\mathrm{base},\mathrm{base},\mathrm{base})$ share
  profile $(0,0)$ but have heights zero and one.
\end{proof}

\begin{theorem}[Orientation Boundary on actual free-recursor successor steps]\label{thm:orientation-boundary-actual}
  Let $e$ be any expression in the reflected scalar grammar of Theorem~\ref{thm:grammar-closure-literal}, and define $M_e(t):=e(c(t),p(t))$ from the observations of Theorem~\ref{thm:free-profile-attainment}. Then
  \[
    \bigl(\forall b,s,n,\;
      M_e(\mathrm{wrap}(s,\mathrm{recur}(b,s,n)))
    < M_e(\mathrm{recur}(b,s,\mathrm{succ}(n)))\bigr)
  \]
  holds if and only if $e$ is payload-independent and counter-strict. If $e$ is counter-strict, the same orientation condition holds if and only if $e$ is payload-independent.
\end{theorem}
\begin{proof}
  Theorem~\ref{thm:free-profile-attainment} identifies uniform orientation of the actual successor instances with the numeric duplicating-step condition. Theorem~\ref{thm:grammar-closure-literal} characterizes that condition as payload blindness plus counter strictness; its counter-strict corollary removes the second conjunct.
\end{proof}

The theorem is an identity between two sets of grammar expressions. Let $\mathcal O$ be the set of expressions whose measure decreases across every actual root successor step of the free recursor, and let $\mathcal B$ be the set of payload-independent, counter-strict expressions. Theorem~\ref{thm:orientation-boundary-actual} states $\mathcal O=\mathcal B$: one side is defined by the rewrite steps, the other by two conditions on the expression alone. This set is the Orientation Boundary. Its complement is inhabited, by the payload projection and by every constant, and the set itself is inhabited, by the counter projection. At every measure and pair one growth cell of Definition~\ref{def:growth-cells} holds, and membership in the barrier cell alone decides failure. Section~\ref{sec:formalization} records the set as a formal object of the Lean development.

The observations above are defined on terms; $p$ counts the designated
wrapper structure. The least
height concerns root rewriting. A successor step stops at the emitted
wrapper under root rewriting, while contextual rewriting continues inside it.

\begin{theorem}[Vector closure under grammar-expressible scalarization]\label{thm:vector-grammar-closure}
  The closure of Theorem~\ref{thm:grammar-closure-literal} extends from scalar measures to vector-valued measures of every finite dimension when the ambient relation is dominated by a scalar functional and the composite of that functional with the vector measure agrees with a grammar expression. The structural premises are domination and grammar agreement. The theorem consists of three strengthenings.

  First, nonincrease suffices. Replacing strict decrease by the weaker requirement that the measure fails to increase across the duplicating step still forces payload independence over the whole grammar. A lexicographic relation forces this condition on its first differing coordinate. The primary-first relation supplies a related diagnostic implication. The theorem proves that relation non-well-founded for every dimension at least three and every designated primary coordinate.

  Second, membership in the grammar replaces the certificate. Call an ambient order $R$ on $\mathbb{N}^d$ \emph{dominated} by a scalar functional $\pi$ when every strict $R$-comparison forces $\pi$ to fail to increase. Let $M$ be a vector measure whose coordinates are grammar expressions, and suppose $\pi\circ M$ agrees with some grammar expression $e$. If $M$ orients the duplicating step under $R$, then $e$ is payload-independent. The mixed-matrix theorem of Section~\ref{sec:barrier} (Theorem~\ref{thm:matrix-arbitrary-scalar-dominance}) asks for a scalarization certificate; here the single requirement is that the scalarization is a grammar expression.

  Third, the standard scalarizations are instances of the general premise. Strict componentwise order is dominated by each coordinate, so orientation forces \emph{every} coordinate to be payload-independent. The primary-coordinate diagnostic forces its designated coordinate to be payload-independent for every coordinate choice and dimension. Every natural-weighted sum of coordinates is itself a grammar expression, obtained by folding scalar multiples and sums, so weighted scalar projections, fixed-row readings, and row sums are covered once the stated domination and grammar-agreement premises are discharged.

  Theorem~\ref{thm:grammar-closure-literal} gives the scalar biconditional over counter-strict grammar expressions: a direct measure orients the step if and only if it is payload-independent. Over the full scalar grammar, counter strictness remains the second conjunct. The vector theorem supplies the corresponding payload independence necessity under its explicit domination and grammar-agreement premises.
\end{theorem}

\noindent The matrix theorems of \S\ref{sec:barrier}
(Theorems~\ref{thm:matrix2-barrier} to~\ref{thm:matrix-arbitrary-scalar-dominance})
name the growth term, its violating triple, and the coefficient
conditions. Theorem~\ref{thm:vector-grammar-closure} applies when its
domination and grammar-agreement conditions hold. An operation that caps a value from above lies outside this grammar.

\smallskip\noindent The last result of this section compares the root rule with its closure under constructor contexts. In one direction the identity context makes every contextual orienter a root orienter. The other direction needs a monotonicity law, and a control example shows that the law is used.

\begin{theorem}[Context lift]\label{thm:context-lift}
  Let $\Rightarrow_c$ be the closure of the two free root rules under one-hole constructor contexts.
  \begin{enumerate}
    \item Every measure that decreases along $\Rightarrow_c$ decreases along both root rules. A measure in the barrier cell at some pair therefore fails to decrease along $\Rightarrow_c$.
    \item Let $I$ interpret the four constructors in a set with a relation $<$, orient both root rules, and satisfy the strict context laws: each operation is strictly increasing in each argument for the relation $<$. Then every $\Rightarrow_c$ step decreases the value of $I$ under every assignment, and $\Rightarrow_c$ terminates when $<$ is well founded.
    \item The interpretation with $\mathrm{wrap}(x,y)=\min(x,y)$, $\mathrm{recur}(b,s,n)=b+n+1$, and $\mathrm{succ}(x)=x+1$ orients both root rules and fails on the contextual step at $\mathrm{wrap}(\mathrm{recur}(\mathrm{zero},\mathrm{zero},\mathrm{succ}(\mathrm{zero})),\mathrm{zero})$, so the conclusion of (ii) uses the strict context laws.
  \end{enumerate}
\end{theorem}

\section{Direct barriers and escape mechanisms}
  \label{sec:barrier}
  The cost/gain theorem becomes a family of impossibility results once a method fixes how the wrapper and successor are interpreted. The barriers use three assumptions: the step payload and the counter are independent arguments, the rule copies that payload into a wrapper, and the direct measure either assigns the wrapper a positive cost or retains it through a tracked scalar. This section records the representative barriers, the classification of tropical interpretations, the theorem that every consumed premise is necessary, and the three escape mechanisms; the Lean development states every family with its violating triple.

  \subsection{Scalar barriers}

  \begin{theorem}[Barrier at the schema]\label{thm:schema-barrier}
    Let $(T,\mathrm{base},\mathrm{succ},\mathrm{wrap},\mathrm{recur})$ be a step-duplicating schema. Then every additive measure with positive wrapper weight fails uniform strict orientation of the duplicating step. The same conclusion holds for every compositional measure whose wrapper strictly dominates each argument and whose successor is transparent at the base point. Hence neither class globally orients any relation containing the duplicating step.
  \end{theorem}
  \begin{proof}
    For an additive measure, freeze $b=n=\mathrm{base}$ and choose a payload whose value reaches the successor contribution. The target then contains the recursive value plus an additional positive wrapper contribution. For the transparent-compositional class, take $b=s=n=\mathrm{base}$: successor transparency makes the recursor values agree before wrapping, while wrapper-subterm strictness makes the target larger. Both arguments yield explicit violating instances.
  \end{proof}

  \begin{theorem}[Affine barrier]\label{thm:affine-barrier}
    Let $M:T\to\mathbb N$ satisfy constructor-local affine equations
    \[
      M(\mathrm{succ}(t))=\alpha_s+\beta_sM(t),\qquad
      M(\mathrm{wrap}(x,y))=\alpha_w+\beta_wM(x)+\gamma_wM(y),
    \]
    \[
      M(\mathrm{recur}(b,s,n))=\alpha_r+\beta_rM(b)+\gamma_rM(s)+\delta_rM(n),
    \]
    with $\beta_w,\gamma_w\ge1$. Then $M$ fails uniform strict orientation of the duplicating step.
  \end{theorem}
  \begin{proof}
    The wrapper coefficients make the target grow at least linearly with the payload, which ranges over $T$, whereas the one-step counter gain is fixed after $b$ and $n$ are frozen. Theorem~\ref{thm:four-cells} therefore places every such interpretation in the unbounded-cost/bounded-gain barrier cell.
  \end{proof}

  The same freeze-and-pump mechanism applies to several nonlinear direct extensions. We state the general bounded-polynomial form because it subsumes the restricted quadratic and multilinear cases used later.

  \begin{theorem}[Bounded-polynomial barrier]\label{thm:polynomial-general-barrier}
    Assume $\mathrm{succ}$ and $\mathrm{wrap}$ are constructor-local affine functions and $\mathrm{recur}$ is a finite natural-coefficient polynomial in $M(b),M(s),M(n)$ of bounded degree. Freeze $b=\mathrm{base}$ and compare the source counter $\mathrm{succ}(\mathrm{base})$ with the target counter $\mathrm{base}$. If the resulting target-side polynomial eventually dominates the source-side polynomial as the payload value grows, then $M$ fails uniform strict orientation. Hence every successful orienter in this family violates that frozen base-dominance condition.
  \end{theorem}

  \subsection{Vector, matrix, and symbolic barriers}

  \begin{theorem}[Tracked componentwise barrier]\label{thm:matrix2-barrier}
    Let $M:T\to\mathbb N^d$ be constructor-local and ordered by strict componentwise decrease. If one coordinate satisfies the affine equations of Theorem~\ref{thm:affine-barrier} with positive wrapper coefficients, then $M$ fails strict componentwise orientation of the duplicating step.
  \end{theorem}

  \begin{theorem}[Mixed-matrix scalarization barrier]\label{thm:matrix-arbitrary-scalar-dominance}
    Let $M:T\to\mathbb N^d$ be a finite-dimensional matrix interpretation. Suppose a nonzero monotone scalar functional $\lambda$ turns every constructor action into the affine scalar interface of Theorem~\ref{thm:affine-barrier}, and suppose every strict comparison in the ambient order forces strict decrease of $\lambda$. Then $M$ fails uniform orientation of the duplicating step.
  \end{theorem}

  \begin{theorem}[Natural-matrix barrier]\label{thm:natural-matrix-barrier}
    Let $M:T\to\mathbb N^d$ be a natural matrix interpretation in finite dimension. If some coordinate $i$ has positive diagonal contribution from both wrapper argument matrices, then no ambient order whose strict comparison forces strict decrease of coordinate $i$ can orient the duplicating step. If strict comparison only forces nonincrease of coordinate $i$, the same conclusion holds once that coordinate takes a positive value on some term.
  \end{theorem}

  \noindent The statement is uniform in the dimension and needs no scalarization certificate, so it is the analogue for the duplicating recursor of the dimension-uniform natural-matrix result of Yolcu, Aaronson, and Heule for the Collatz system~\cite{YolcuAaronsonHeule23}. Over the tropical semiring the root rule has an orienting interpretation on free recursor syntax, in dimension one, and the rule closed under both wrapper contexts has none in any finite dimension.

  \begin{theorem}[Tropical root and context classification]\label{thm:tropical-context-barrier}
    Let $M$ be a finite-dimensional tropical (min-plus) natural matrix interpretation of a step-duplicating schema and fix a coordinate $i$. Write $\Rightarrow_{\mathrm{dup}}$ for the closure of the duplicating rule under both wrapper arguments. Then no such $M$ satisfies $M(y)_i<_{\mathrm{trop}}M(x)_i$ for every $x\Rightarrow_{\mathrm{dup}}y$, with no hypothesis on the wrapper diagonals, biases, attained values, or scalarization, and the conclusion holds for every relation containing $\Rightarrow_{\mathrm{dup}}$. On free recursor syntax a one-dimensional tropical interpretation orients every duplicating root instance.
  \end{theorem}

  \begin{corollary}[Wrapper-congruence policies]\label{cor:tropical-wrapper-policy}
    On free recursor syntax, close the duplicating rule under any subset of the two wrapper-congruence rules. A finite-dimensional tropical natural matrix interpretation strictly decreases a fixed coordinate on this relation if and only if at least one wrapper-congruence rule is absent, and dimension one suffices in every positive case.
  \end{corollary}

  \noindent Under the right wrapper congruence, strict orientation of $\mathrm{wrap}(\mathrm{base},u)\Rightarrow\mathrm{wrap}(\mathrm{base},v)$ for a root step $u\to v$ forces the right contribution at $v$ to be finite, say $N$; every term $\mathrm{wrap}(x,v)$ then has tracked value at most $N$, and an execution of $N+1$ contractions placed in the left argument contradicts strict descent below that bound. The positive cases use the interpretations that read only the left, or only the right, wrapper argument.

  \begin{theorem}[Variable-condition barrier]\label{thm:symbolic-variable-barrier}
    Let $\succ$ be any strict symbolic comparator satisfying the standard variable condition: $l\succ r$ implies that every variable occurs in $r$ at most as often as in $l$. Then $\succ$ cannot orient
    \[
      \mathrm{recur}(b,s,\mathrm{succ}(n))\to\mathrm{wrap}(s,\mathrm{recur}(b,s,n))
    \]
    from left to right.
  \end{theorem}

  \begin{corollary}[KBO impossibility]\label{cor:kbo-impossible}
    Every KBO-style comparator satisfying the variable condition fails on the concrete successor recursor rule, because the step variable occurs once on the source side and twice on the target side.
  \end{corollary}

  \begin{theorem}[Subterm-coefficient KBO barrier]\label{thm:subterm-coefficient-kbo}
    In a Knuth-Bendix order with subterm coefficients, every argument position $i$ of a symbol $f$ carries a coefficient $c_i(f)\ge1$, and the variable condition is read on the coefficient-weighted occurrence count $\mathrm{vc}$, with $\mathrm{vc}(x,x)=1$, $\mathrm{vc}(x,y)=0$ for $y\ne x$, and $\mathrm{vc}(x,f(t_1,\dots,t_n))=\sum_i c_i(f)\,\mathrm{vc}(x,t_i)$. On the duplicating rule the payload variable has weighted count $c_2(\mathrm{recur})$ on the source side and $c_1(\mathrm{wrap})+c_2(\mathrm{wrap})\,c_2(\mathrm{recur})\ge1+c_2(\mathrm{recur})$ on the target side, so no such order orients the step for any coefficient assignment. Each positivity assumption is used: setting either wrapper coefficient to zero removes the obstruction.
  \end{theorem}

  \noindent Orientability by Knuth-Bendix orders with subterm coefficients is undecidable over all input systems~\cite{MitterwallnerMiddeldorpThiemann24}; on the duplicating rule the answer is fixed before weights or precedence are considered.

  \subsection{Every consumed premise is necessary}

  Each barrier theorem has stated premises. The next theorem records, for every premise a proof consumes, a measure that satisfies the other premises, violates that one, and orients the duplicating rule, and for every stated premise outside the proof, the theorem with that premise removed.

  \begin{theorem}[Premise necessity for the direct families]\label{thm:premise-necessity}
    Across the twelve direct families of Corollary~\ref{cor:families-barrier-cell}, twenty premises are consumed by the barrier proofs. For each of them there is a measure of the family, on the free schema, that satisfies every other premise of its theorem, violates that premise, and orients the duplicating rule at every triple; the barrier with that premise deleted is false. The three families with no consumed premise (max-plus, maximum depth, head precedence) have barriers that hold from the family equations alone. The wrapper positivity premises of the cross-term, multilinear, and polynomial theorems, the first wrapper-subterm axiom of the compositional theorem, the wrapper weight of the additive theorem, and the right offset $\gamma_w\ge1$ of the max-plus theorem are stated and unused: each of those theorems holds with it removed.
  \end{theorem}

  \noindent Representative countermodels: the counter projection and the payload projection, read as affine measures, delete the two wrapper positivity premises of Theorem~\ref{thm:affine-barrier}; the zero measure deletes the positive-value premise of the lexicographic barrier; the coupled interpretation $M(\mathrm{recur}(b,s,n))=M(n)+M(s)M(n)$ deletes the dominance premise of the cross-term, multilinear, and polynomial theorems at once; the negative integers delete the attainment premise of the ordered-carrier additive barrier. The Lean development records all twenty as premise slots, each with its countermodel.

  \subsection{Escape mechanisms}

  The barrier concerns direct orientation. Three positive constructions show how a proof succeeds after changing the information available to the comparison.

  \begin{theorem}[Nonlinear direct escape]\label{thm:transparency-essential}
    There is a compositional natural-valued measure satisfying both wrapper-subterm inequalities but not successor transparency that strictly orients every duplicating successor instance. One example is
    \[
      c_{\mathrm{base}}=1,\quad c_{\mathrm{succ}}(x)=x+1,\quad
      c_{\mathrm{wrap}}(x,y)=x+y+1,
    \]
    \[
      c_{\mathrm{recur}}(a,b,c)=a+b(c+1)^2+c.
    \]
    Thus the transparent-compositional barrier uses its successor-transparency hypothesis.
  \end{theorem}

  \begin{theorem}[Free recursor normalization]\label{thm:free-source-normalization}
    For the free two-rule recursor, both root rewriting and full contextual rewriting are terminating and confluent. Every term has a computable normal form $N(t)$, and two terms are joinable if and only if their normal forms agree.
  \end{theorem}

  \begin{theorem}[Dependency-pair projection]\label{thm:dp-escape}
    Define $\pi$ on the free recursor by
    \[
      \pi(\mathrm{base})=0,\qquad \pi(\mathrm{succ}(t))=\pi(t)+1,
      \qquad \pi(\mathrm{wrap}(x,y))=0,
      \qquad \pi(\mathrm{recur}(b,s,n))=\pi(n).
    \]
    Then the transformed recursive call
    \[
      \mathrm{recur}(b,s,\mathrm{succ}(n))\to_{\mathrm{DP}}\mathrm{recur}(b,s,n)
    \]
    strictly decreases $\pi$ by one. The projection erases the duplicated payload and therefore violates both wrapper-sensitivity assumptions used by the direct barrier.
  \end{theorem}

  \begin{theorem}[Dependency-pair normalization]\label{thm:dp-normalization}
    The transformed-call relation is deterministic, terminating, and confluent. Its normalizer removes the leading successor prefix of the counter, and every path to normal form has length equal to that prefix. The same statement holds for the corresponding CWS pair relation.
  \end{theorem}

  \begin{proposition}[A linear base order already suffices after transformation]\label{prop:dp-base-order-boundary}
    The extracted CWS pair problem admits a linear natural-valued base order that strictly decreases on every pair. Dependency-pair extraction therefore removes the direct affine barrier: an affine order orients the transformed problem.
  \end{proposition}

  A second escape imports the subterm property and keeps the payload information.

  \begin{proposition}[Fixed-signature path order]\label{prop:lpo-full-context}
    For the fixed CWS signature and precedence, the lexicographic path order with full lexicographic status is transitive, stable under substitution, monotone under every finite one-hole context, and reverse well founded. Every root and full-context CWS step decreases in this order.
  \end{proposition}

  \begin{proposition}[Nonlinear polynomial full-step escape]\label{prop:poly-full-step}
    There exists a natural-valued interpretation $W$ that strictly decreases on every CWS root step. It couples the counter and payload multiplicatively and is not successor-transparent, so it lies outside the additive, transparent-compositional, and affine barrier classes.
  \end{proposition}

  \begin{proposition}[Specialized MPO termination]\label{prop:mpo-full-step}
    For the fixed CWS signature, precedence, and same-head clause, the reverse root relation induced by the CWS rules is well founded. The same-head recursor clause compares the counter argument while requiring equality of the base and step arguments.
  \end{proposition}

  \begin{proposition}[Ordinal bound for the specialized MPO proof]\label{prop:mpo-proof-theoretic-bound}
    The specialized MPO proof admits an ordinal ranking below $\varphi_7(0)<\Gamma_0$.
  \end{proposition}

  \begin{theorem}[Escape trichotomy]\label{thm:escape-trichotomy}
    Let $\mathcal O$ be a successful direct root-step orienter in the explicit direct universe considered above, and let $\pi_{\mathcal O}$ be its tracked scalar. At least one of the following holds:
    \begin{enumerate}
      \item $\pi_{\mathcal O}$ loses wrapper sensitivity;
      \item $\pi_{\mathcal O}$ loses successor transparency at the base term;
      \item $\mathcal O$ leaves the explicit direct universe of additive, transparent-compositional, affine, bounded-polynomial, max-plus, tracked vector/matrix, max-depth, and pure head-precedence families.
    \end{enumerate}
  \end{theorem}

  \begin{proposition}[Additive-sum Yamada tuple fragment]\label{prop:yamada-additive-sum}
    The coordinate-sum of every additive-sum Yamada tuple interpretation is an additive scalar measure, so the additive barrier applies to the duplicating schema step. No interpretation of the affine strict-coordinate fragment with positive wrapper coefficients orients the duplicating step (Lean theorem \path{no_yamadaAffineTuple_orients_dup_step}).
  \end{proposition}
  \noindent The summed evaluation of a Yamada tuple measure is an additive measure. An orientation of the duplicating step in the tuple order would be an orientation of the summed evaluations, which the additive barrier excludes. \hfill$\square$

  \noindent Sharing is an escape that changes the representation. Once the two copies of the payload share one counter, a counter measure orients the shared step, so the barrier is a theorem about independent copies; a directed collapse rule $\mathrm{wrap}(t,t)\to t$, by contrast, keeps every direct barrier because the duplicating rule is unchanged. Removing duplication itself removes the barrier: the linear variant $\mathrm{recL}(b,s,\mathrm{succ}(n))\to\mathrm{recL}(b,s,n)$ is oriented by term size. The Lean development proves these ablations, the arctic classification, the alternating-cycle and multiplicity-preserving extensions, and every remaining family with its violating triple.

\subsection{Seventy-six termination methods on the free recursor}\label{sec:method-classification}

The barrier theorems and the escapes are stated by measure class. The same question can be asked method by method. Table~\ref{tab:method-groups} groups seventy-six named termination methods from the literature by the mechanism that decides their termination verdict on the free recursor, and Appendix~\ref{sec:methods} lists every method with its citation.

\begin{theorem}[Method classification]\label{thm:method-classification}
  For each of the seventy-six method families of Table~\ref{tab:method-groups} there are typed method data, the laws of the method, and a proof of the stated termination verdict: a barrier, an escape, or the named termination or equivalence result. Every row has a necessity statement. Fourteen rows use a barrier premise whose deletion admits an orienting countermodel; seven reject the rule by an obstruction that uses no barrier premise and come with a control example; forty-two change termination verdict when the escape datum, a projection, labelling, sort assignment, or order, is changed; thirteen have a termination verdict fixed by their defining laws. Thirteen rows whose method is a named theorem of the literature contain that theorem in the form its source states it.
\end{theorem}

\begin{table}[ht]
  \centering\small
  \caption{Seventy-six termination methods on the free recursor, grouped by the mechanism that decides the termination verdict.}
  \label{tab:method-groups}
  \begin{tabularx}{\textwidth}{@{}P{4.4cm}cY@{}}
    \toprule
    Deciding mechanism & Rows & Methods and termination verdict \\
    \midrule
    Barrier cell (Theorem~\ref{thm:four-cells}) & 11 & Knuth-Bendix orders (standard, with subterm coefficients, with status, transfinite, lambda-free, AC-compatible), matrix interpretations over $\mathbb N$, $\mathbb Q$, $\mathbb R$, triangular matrices: barrier \\
    Order of the carrier & 3 & arctic and tropical interpretations, strictly monotone algebras on a finite carrier: barrier \\
    Derivation length (Theorem~\ref{thm:free-exponential}) & 2 & match bounds, match bounds for left-linear systems: barrier \\
    Syntactic condition of the method & 5 & right-forward closures, string rewriting, cycle rewriting, type introduction: barrier; linear typing without exponentials: fragment barrier \\
    Coupling counter gain to the payload (Theorem~\ref{thm:context-lift}) & 10 & polynomial interpretations of any degree, negative coefficients, max-polynomials, multilinear interpretations, monotone and extended monotone algebras, polynomial-weight KBO: escape; higher-order tuple interpretations and generalized KBO: the named termination result; additive-sum tuple fragment: barrier \\
    Projection to the counter (Theorem~\ref{thm:dp-escape}) & 18 & subterm criterion, argument filtering, reduction pairs and triples, neutral processors and processor sequences, size-change termination, usable and formative rules, context-sensitive, order-sorted, conditional, integer, and constrained dependency pairs: escape; persistence under sorts: all-term termination from sorted termination \\
    Structural recursion on the counter & 5 & guard condition, safe recursion, ramified recurrence, sized types, general schema: fragment escape \\
    Precedence, subterm, and ordinal descent & 8 & recursive path order with status, polynomial path order, AC-compatible RPO, slow-growing bound, quasi-decreasingness: escape; HORPO and computability path order: fragment escape; simplification orders: RPO soundness over every signature \\
    Labelling & 5 & semantic, predictive, root, and self-labelling, match bounds with raise consistency: escape \\
    Changing the representation & 9 & rewriting modulo equations, term graph rewriting with sharing: nonconservative escape; weighted type graphs, process-calculus and CPS translations, quasi-interpretations with sharing, well-founded coalgebras, infinitary rewriting: the named termination or equivalence result \\
    \bottomrule
  \end{tabularx}
\end{table}

\section{A decision theorem for polynomial orientation}
\label{sec:decision}

Polynomial interpretations form the class in which the criterion of Section~\ref{sec:criterion} becomes an algorithm. A \emph{constructor-local natural polynomial interpretation} of the free schema sends $\mathrm{zero}$ to a number $z$, $\mathrm{succ}(n)$ to $\beta n+a$, $\mathrm{wrap}(s,y)$ to a polynomial $W(s,y)$, and $\mathrm{recur}(b,s,n)$ to a polynomial $R(b,s,n)$, all with natural coefficients. It orients a rule when the value of the left side exceeds the value of the right side under every assignment of natural numbers to the variables. Mitterwallner, Middeldorp, and Thiemann proved that orientability by linear polynomial interpretations over the natural numbers, and by Knuth-Bendix orders with subterm coefficients, is undecidable when the input is an arbitrary finite rewrite system~\cite{MitterwallnerMiddeldorpThiemann24,MitterwallnerMiddeldorpThiemannAFP24}. The theorems below fix the input system, the two rules of the free recursor, and let the interpretation range over polynomials of every degree.

\begin{theorem}[Decision for unit successor]\label{thm:tdec1}
  A total procedure takes $z,a\in\mathbb{N}$ and two lists of monomials $W$ and $R$ and returns \emph{true} if and only if the interpretation with successor $n\mapsto n+a$ orients both root rules of the free recursor. When it rejects the successor rule, a second procedure returns a triple $(b,s,n)$ with $R(b,s,n+a)\le W(s,R(b,s,n))$. The same equivalence holds over every first-order signature that contains the four schema symbols, for every interpretation of the remaining symbols.
\end{theorem}

\noindent Three syntactic tests exclude orientation of the successor rule for every slope $\beta\ge1$: a wrapper monomial of $y$-degree at least two; a $y$-coefficient of at least $\beta^d+1$ at some payload, where $d$ is the largest counter degree of $R$; and a wrapper monomial $c\,s^j y$ with $c,j\ge1$, which reaches that coefficient at $s=\beta^d+1$. Each test returns its failing triple. On the remaining wrappers, $W(s,y)=w\,y+C(s)$, and for $w\le1$ orientation is the comparison of two polynomials in the payload alone. The comparison $P(x)<Q(x)$ for every natural $x$ is decided by testing the finitely many values $x\le P(1)+Q(1)+1$. With $\beta=1$ the leading-coefficient test already excludes every wrapper with $w\ge2$, so the procedure returns an answer on every input, and the zero rule reduces to a comparison of the same kind. \hfill$\square$

\begin{corollary}[Two classes decided]\label{cor:region-decides}
  For $W(s,y)=s+y+1$ and $R(b,s,n)=(n+1)(\alpha s+b+\beta)$ with $z=0$ and $a=1$, the procedure of Theorem~\ref{thm:tdec1} accepts if and only if $\alpha\ge1$ and $\beta\ge2$. For an affine wrapper $W(s,y)=\alpha+\beta_w s+\gamma y$ over the multilinear recursor
  \[
    R(b,s,n)=r_0+r_b b+r_s s+r_n n+r_{bs}bs+r_{bn}bn+r_{sn}sn+r_{bsn}bsn
  \]
  with successor $n\mapsto n+a$, the successor rule is oriented if and only if either $\gamma=0$, $\alpha<r_0+r_n a$, and $\beta_w\le r_s+r_{sn}a$, or $\gamma=1$, $\alpha<r_n a$, and $\beta_w\le r_{sn}a$. In particular every affine interpretation that keeps both wrapper arguments, $\beta_w\ge1$ and $\gamma\ge1$, over an affine recursor is rejected.
\end{corollary}

\noindent The first statement obtains the polynomial region of Theorem~\ref{thm:free-source-normalization} as an output of the decision. The second applies the coupling inequality inside one class: the gain of a multilinear recursor grows with the payload only through the monomials $sn$ and $bsn$, and the answer records whether that growth is at least the value of the retained copy of $s$.

\begin{theorem}[Affine successor: the decided class and its residual]\label{thm:tdec2}
  For every slope $\beta\ge1$, a procedure returns, for the successor rule under $n\mapsto\beta n+a$, either a correct answer together with a failing triple on rejection, or no answer. It returns no answer if and only if $W$ has no monomial of $y$-degree at least two and no monomial $c\,s^j y$ with $c,j\ge1$, the $y$-coefficient $w$ of $W$ satisfies $2\le w\le\beta^d$, and some monomial of $R$ with positive coefficient has a counter degree $k$ with $k=0$ or $w>\beta^k$. For $\beta=1$ it always returns an answer. For $\beta=2$ the class without an answer contains an interpretation that orients the successor rule and one that fails to orient it. On that class, orientation holds if and only if
  \[
    \sum_{m} c_m\,b^{i_m}s^{j_m}\bigl((\beta n+a)^{k_m}-w\,n^{k_m}\bigr)-W(s,0)>0
    \qquad\text{for all }b,s,n\in\mathbb{N},
  \]
  where the sum runs over the monomials $c_m b^{i_m}s^{j_m}n^{k_m}$ of $R$.
\end{theorem}

\noindent For slopes $\beta\ge2$, Theorem~\ref{thm:tdec2} isolates the undecided remainder of this procedure as one explicit signed positivity problem in $(b,s,n)$. Theorem~\ref{thm:residual-bivariate} shows that this remainder contains every signed bivariate polynomial difference, and Corollary~\ref{cor:residual-zero-reduction} states the reduction: a total decision for the whole residual would decide zero existence for arbitrary integer polynomials over $\mathbb N^2$. The positive higher-slope results are the six named decidable strata of Theorem~\ref{thm:residual-strata}.

\begin{theorem}[Two-variable polynomials inside the unanswered class]\label{thm:residual-bivariate}
  Let $P^{+}$ and $P^{-}$ be polynomials in $b$ and $s$ with natural coefficients, and set
  \[
    W(s,y)=2y,\qquad R_P(b,s,n)=n\bigl(1+P^{+}(b,s)\bigr)+1+P^{-}(b,s),\qquad \beta=2,\quad a=1.
  \]
  The procedure of Theorem~\ref{thm:tdec2} leaves $(W,R_P)$ unanswered, and the positivity form of Theorem~\ref{thm:tdec2} equals $P^{+}(b,s)-P^{-}(b,s)$ at every $b,s,n\in\mathbb{N}$. Hence $R_P$ orients the successor rule if and only if $P^{+}(b,s)>P^{-}(b,s)$ for all $b,s\in\mathbb{N}$. Every integer polynomial in $b$ and $s$ is such a difference $P^{+}-P^{-}$, and the interpretation built from the difference representing $(P^{+}-P^{-})^{2}$ orients the successor rule if and only if $P^{+}(b,s)\neq P^{-}(b,s)$ for all $b,s\in\mathbb{N}$.
\end{theorem}

\noindent For a monomial of counter degree one the bracket $(2n+1)-2n$ equals $1$; for counter degree zero it equals $-1$. The monomial $n$ of $R_P$ contributes $+1$, the constant $1$ contributes $-1$, and $W(s,0)=0$, so the form reduces to $P^{+}-P^{-}$ for every counter value. The wrapper has $y$-coefficient $2=\beta^{d}$ with $d=1$, and the constant monomial of $R_P$ has counter degree zero, so the input satisfies the description of the unanswered class. For the square, $x^{2}>0$ holds if and only if $x\neq0$. \hfill$\square$

\begin{corollary}[A decision for the unanswered class decides two-variable zeros]\label{cor:residual-zero-reduction}
  Suppose a total procedure takes an input of the unanswered class of Theorem~\ref{thm:tdec2} with slope $\beta\ge2$, returns \emph{true} if and only if the positivity form is positive for all $b,s,n\in\mathbb{N}$, and returns a failing triple otherwise. Applying it to the square construction of Theorem~\ref{thm:residual-bivariate} and negating the answer gives a total procedure that decides, for every integer polynomial in two variables, whether it has a zero in $\mathbb{N}^{2}$.
\end{corollary}

\noindent The corollary places the whole residual beside Hilbert's tenth problem: the zero problem for integer polynomials over the natural numbers is undecidable, and undecidable already in nine unknowns~\cite{Davis73,Jones82}, while the two-variable case is open over the integers~\cite{Poonen08}. A total decision for the whole residual would settle that case, because an integer polynomial $P$ in two variables has a zero in $\mathbb Z^2$ if and only if $P(x,y)\,P(-x,y)\,P(x,-y)\,P(-x,-y)$ has a zero $(x,y)\in\mathbb N^2$. The remainder of this section identifies six named subfamilies on which positivity is decided by total procedures.

\begin{theorem}[Six named decidable strata inside the higher-slope residual]\label{thm:residual-strata}
  On each of the following classes, a total procedure decides whether the positivity form of Theorem~\ref{thm:tdec2} is positive for all $b,s,n\in\mathbb{N}$ and returns a failing assignment on rejection. Membership is computable from the monomial lists.
  \begin{enumerate}
    \item \emph{Coefficient-monotone inputs:} $w\le\beta$, and every monomial of $R$ with a positive coefficient has counter degree one, or has counter degree zero and base degree zero. The form is nondecreasing in $b$ and $n$, so positivity reduces to a univariate comparison in $s$ at $b=n=0$.
    \item \emph{Base-affine inputs:} for the residual encodings $R_P$ of Theorem~\ref{thm:residual-bivariate}, $P^{+}$ and $P^{-}$ have degree at most one in $b$. Writing $P^{+}-P^{-}=A(s)+bB(s)$, positivity is equivalent to $A(s)>0$ and $B(s)\ge0$ for every $s$.
    \item \emph{Payload-affine inputs:} the symmetric class in which $P^{+}$ and $P^{-}$ have degree at most one in $s$; the same criterion holds with $b$ and $s$ exchanged.
    \item \emph{Payload-only inputs:} every monomial has base degree zero. Positivity is decided by the resulting univariate polynomial in $s$, and rejection returns a failing payload coordinate.
    \item \emph{Base-only inputs:} every monomial has payload degree zero. Positivity is decided by the resulting univariate polynomial in $b$, and rejection returns a failing base coordinate.
    \item \emph{Monomial-difference inputs:} for the residual encodings $R_P$, $P^{+}$ and $P^{-}$ each consist of one monomial. The procedure accepts when the positive monomial is a constant of coefficient at least one and the negative monomial has coefficient zero or is a constant whose coefficient is smaller. Rejection returns a failing pair $(b,s)$.
  \end{enumerate}
  The payload-only class is contained in the base-affine class and the base-only class is contained in the payload-affine class. We give direct decision procedures and failure extractors for both univariate strata because they yield simpler certificates and counterexamples.
\end{theorem}

\noindent In family 2, a failing triple comes from the first payload $s_0$ where a univariate comparison fails: if $A(s_0)\le0$ the triple is $(0,s_0,0)$; if $A>0$ everywhere and $B(s_0)\le-1$, the triple is $(b_0,s_0,0)$, where $b_0=A^{+}(s_0)$ is the value of the base-free part of $P^{+}$; since $A(s_0)\le A^{+}(s_0)$, the value $A(s_0)+b_0B(s_0)$ is at most $0$. The polynomial $3b^{2}s-5b+7$ lies in family 3 and is rejected. The polynomial $b^{2}s^{2}-5b$ belongs to family 6; families 1, 2, and 3 exclude it; the procedure rejects it at $(0,0)$. \hfill$\square$

\smallskip\noindent\textbf{Position beside the undecidability theorem.} The two results answer different questions. The theorem of Mitterwallner, Middeldorp, and Thiemann quantifies over input systems and concerns linear interpretations. Theorem~\ref{thm:tdec1} fixes one input system and decides orientation for polynomial interpretations of every degree with unit successor. Neither statement implies the other. For linear interpretations of the free recursor, Corollary~\ref{cor:region-decides} gives the answer: every interpretation that keeps both wrapper arguments is rejected.

\smallskip\noindent The next result states the boundary as a property of checkable certificates. A certificate is a polynomial interpretation, a natural matrix interpretation with positive upper-left entries in the sense of Endrullis, Waldmann, and Zantema~\cite{EndrullisWaldmannZantema08}, a Knuth-Bendix weight and precedence assignment, or a dependency-pair projection, and each kind has an executable checker.

\begin{theorem}[The boundary in a certificate language]\label{thm:certificate-boundary}
  Every certificate accepted by its checker proves termination of the free recursor. The checkers reject every affine polynomial certificate that keeps both wrapper arguments, every matrix certificate, and every Knuth-Bendix certificate. They accept the dependency-pair certificate that projects the recursive call to its counter and the polynomial certificate $W=s+y+1$, $R=(n+1)(s+b+2)$. On certificates whose monomials are linear in each wrapper and recursor argument, the polynomial checker accepts if and only if the interpretation orients both root rules.
\end{theorem}

\noindent The matrix and Knuth-Bendix rejections are the coupling inequality and the variable condition read on the certificate data: at the tracked entry the step argument occurs once on the left of the successor rule and twice on the right, and positive entries or weights make the right side larger.

\begin{theorem}[Sorted persistence for the free recursor]\label{thm:free-recursor-sorted-persistence}
  Let $S$ be any type of sorts and let $A$ assign result and argument sorts to the free-recursor signature. Suppose that the left- and right-hand sides of each recursor rule have a common sort under $A$. Termination of the sorted step relation on every well-sorted term is equivalent to strong normalization of the unsorted free-recursor relation on every term. A concrete two-sort assignment satisfies the rule-typing and sorted-termination premises. Hence the unsorted relation is strongly normalizing on every term, including ill-sorted terms and applications with an incorrect arity.
\end{theorem}

\begin{proof}
  For a fixed result sort, replace each principal subterm of a different sort by a variable of its active sort. Rule typing supplies variables only for the sorts used by the signature; unused sorts need no inhabitant. A rewrite step either decreases the sorted projection or preserves that projection and decreases the cut-expanded multiset of replaced subterms. Lexicographic well-foundedness therefore transfers accessibility from the sorted relation. An application with an incorrect arity has no root step and inherits accessibility from its arguments. The converse follows because the sorted relation is a subrelation of the unsorted relation. For the concrete two-sort assignment, induction on a well-sorted counter proves sorted termination, so the equivalence gives strong normalization of every free term. Under rule typing and the condition that every right-hand-side variable occurs on the left, the native sorted substitution relation is equivalent to the sorted step relation. This result uses substitution-based many-sorted rewriting. The restriction to a system whose rules the assignment types is used: Toyama's system is terminating on well-sorted terms, has a nonterminating ill-sorted loop, and fails Zantema's condition, so sorting alone transfers no termination for an arbitrary rule list.
\end{proof}

\section{Semantic measures and factorization}
  \label{sec:schema-diagnostic}
  The scalar grammar gives a characterization inside a closed class of direct measures. The next result isolates a semantic condition under which the same information loss is forced for \emph{arbitrary} carrier-valued measures. Its domain is one root step pair.

  \begin{definition}[Payload erasure]\label{def:payload-erasure}
    Let an abstract recursor step have source $\ell(b,s,n)$ and target $r(b,s,n)$. A \emph{payload erasure} consists of a map $e:T\to T$ and a neutral payload $s_0$ such that, for every $b,s,n$,
    \[
      e(\ell(b,s,n))=\ell(b,s_0,n),\qquad
      e(r(b,s,n))=r(b,s_0,n).
    \]
    Thus $e$ removes the payload from both sides and keeps the base and counter data of the step.
  \end{definition}

  \begin{theorem}[Arbitrary semantic payload-erasure theorem]\label{thm:s10-arbitrary-semantic-classifier}
    Let $A$ carry a well-founded strict relation $<$, let $\mu:T\to A$, and suppose $\mu$ orients the recursor step:
    \[
      \mu(r(b,s,n))<\mu(\ell(b,s,n))
      \qquad\text{for all }b,s,n.
    \]
    If the step carries a payload erasure $(e,s_0)$, then the erased measure
    \[
      \mu_0(t):=\mu(e(t))
    \]
    is payload-independent on both sides of the step and also orients every instance. Hence every orienting semantic measure is counter-dominated: some payload-independent measure also orients the step.
  \end{theorem}
  \begin{proof}
    Payload blindness follows from the two erasure equations. For orientation,
    \[
      \mu_0(r(b,s,n))=\mu(r(b,s_0,n))
      <\mu(\ell(b,s_0,n))=\mu_0(\ell(b,s,n)).
    \]
  \end{proof}

  \begin{corollary}[No decisive payload-sensitive certificate under erasure]\label{cor:s10-no-decisive-certificate}
    Under the hypotheses of Theorem~\ref{thm:s10-arbitrary-semantic-classifier}, no orientation certificate can be both payload-sensitive and decisive in the sense that every payload-independent alternative fails: the erased measure $\mu_0$ is such an alternative and still orients.
  \end{corollary}

  The erasure hypothesis is used: the Lean development exhibits a root step without an erasure and an orienting semantic measure on it that is not counter-dominated. The theorem therefore identifies a structural reason for payload irrelevance, a static erasure of the payload, and the free recursor has one.

  \begin{proposition}[Projection-transaction escape]\label{prop:s4-projection-transaction}
    Let $\pi:T\to T'$ commute with the source and target of a step, let $T'$ have a well-founded strict order, and suppose a measure on $T'$ strictly orients the projected step. If $\pi$ factors away the duplicated payload, then the composite measure on $T$ orients the source step through the projection. Each of commutation, well-foundedness, and the factorization property is used in the transport.
  \end{proposition}

  For the free recursor, dependency-pair projection is the standard example: $\pi$ keeps the counter and discards the payload. The transformed recursive call then becomes the strict natural descent $n+1\mapsto n$.

  An observation $q$ of states \emph{licenses a rank} when the relation that the recursive call induces on observed values is well founded in reverse; equivalently, some rank into a well-founded relation factors through $q$. One observation \emph{refines} another when equal values of the first imply equal values of the second.

  \begin{theorem}[Observers that license the recursive-call rank]\label{thm:observer-sufficiency}
    Let $c$ be the counter observation on free recursor calls.
    \begin{enumerate}
      \item An observation $g\circ c$ licenses a rank if and only if $g$ is injective on counter values.
      \item The observations sufficient for the counter are those refining the kernel quotient of $c$; the quotient itself is sufficient.
      \item There are natural-valued observations that license ranks while neither refines the counter nor the other. Thus the counter is least informative only among observations that factor through it.
      \item The natural-valued rankings constant on counter fibers are the strictly increasing recodings of $c$ and induce one strict order. Without the fiber-constancy restriction, infinitely many strict orders occur.
    \end{enumerate}
  \end{theorem}

  \begin{proposition}[Aggregation of recursive arguments]\label{prop:aggregation-projection}
    For direct grammar measures on the $(\mathrm{counter},\mathrm{payload})$ carrier:
    \begin{enumerate}
      \item a scalar measure orients if and only if it is payload-independent and counter-strict;
      \item under lexicographic comparison, orientation forces the first coordinate to be payload-independent, while later coordinates may retain payload information;
      \item under strict componentwise comparison, every coordinate must be payload-independent.
    \end{enumerate}
    Thus the scalar case and the leading lexicographic coordinate factor through projection to the counter.
  \end{proposition}

  \paragraph{Scope.}
  The arbitrary theorem quantifies over every semantic measure satisfying the payload-erasure interface. The finite reflected classifiers of the Lean development give computable classifications for their declared syntactic certificate languages.

\section{Derivation length and ordinal height on the free recursor}
\label{sec:costs}

Theorem~\ref{thm:free-source-normalization} proves that the free recursor terminates under all contexts. This section measures its derivations and the ordinals its termination proofs use. The size $|t|$ of a term counts its constructors and variables.

\begin{theorem}[Exponential derivation length]\label{thm:free-exponential}
  Every contextual derivation of the free recursor from a term $t$ has at most $3^{|t|}-2$ steps. For $k\in\mathbb{N}$ let
  \[
    t^{(k)}_0=\mathrm{recur}(\mathrm{zero},\mathrm{zero},\mathrm{zero}),
    \qquad
    t^{(k)}_{d+1}=\mathrm{recur}(\mathrm{zero},t^{(k)}_d,\mathrm{succ}^k(\mathrm{zero})).
  \]
  Then $t^{(k)}_d$ has a contextual derivation of at least $k^d$ steps. Since $|t^{(2)}_d|=5d+4$, derivations from terms of size $5d+4$ reach length $2^d$.
\end{theorem}

\noindent The upper bound comes from a polynomial interpretation that decreases by at least one along every contextual step and satisfies $w(t)+2\le3^{|t|}$. The lower bound unfolds the outer recursor $k$ times, which leaves $k$ copies of $t^{(k)}_d$ under wrappers, and then reduces every copy: duplication before reduction multiplies the lengths. The dependency-pair escape is linear on its own relation. Along a chain of $m$ recursive calls from $a$ to $c$, the number $m$ plus the counter of $c$ equals the counter of $a$, where the counter of a recursor term is the length of the leading successor prefix of its third argument (Theorem~\ref{thm:dp-normalization} describes the normal forms).

\begin{theorem}[Ordinal height of the escapes]\label{thm:free-ordinal-height}
  On free recursor syntax:
  \begin{enumerate}
    \item The height of every term in the reverse contextual relation is a natural number, every natural number is the height of some term, and the supremum of the heights is $\omega$. The height of every term in the reverse recursive-call relation is a natural number, and every natural number occurs.
    \item The path order generated by the subterm relation, by the precedence $\mathrm{zero}<\mathrm{succ}<\mathrm{wrap}<\mathrm{recur}$ with dominated arguments, and by decrease of the counter under an equal $\mathrm{recur}$ head orients both root rules. An ordinal ranking decreases along this order, every value of the ranking lies below $\varphi_4(0)$ in the Veblen hierarchy, and the supremum of the values is $\varphi_4(0)$.
  \end{enumerate}
\end{theorem}

\noindent The natural-valued escapes, the counter projection and the polynomial interpretation, are therefore calibrated at $\omega$. The CWS-specialized path order of Proposition~\ref{prop:mpo-proof-theoretic-bound} ranks terms over seven constructors below $\varphi_7(0)$; over the four schema symbols the corresponding ranking has supremum $\varphi_4(0)$.

\section{Transport: signatures, arity, and System~T}
\label{sec:transport}

The theorems of Sections~\ref{sec:criterion} and~\ref{sec:decision} quantify over step-duplicating schemas or over the free recursor. This section connects other systems to those objects by maps of a stated kind: an isomorphism of rewrite relations, a map in one direction, or a proof that a map of a given shape fails to exist. For each map it states which other maps of the same kind exist.

\subsection{Arbitrary signatures}

\begin{theorem}[Signature transport]\label{thm:signature-transport}
  Let $\Sigma$ consist of the four schema symbols together with further function symbols of arbitrary arities, and let the two recursor rules be the root rules over $\Sigma$-terms. The inclusion of free recursor syntax into $\Sigma$-terms is injective and commutes with substitution and with one-hole contexts. Its image is the set of terms without additional symbols; the image is closed under root and contextual steps, and every step from an image term is the image of a free step. The barrier cell theorem, the coupling inequality, contextual termination under the strict context laws, and the decision of Theorem~\ref{thm:tdec1} hold over $\Sigma$. The free schema is the instance without additional symbols. CWS is the instance in which $\mathrm{void}$, $\delta$, $\mathrm{app}$, and $\mathrm{rec}\Delta$ are the schema symbols and $\mathrm{integrate}$, $\mathrm{merge}$, and $\mathrm{eqW}$ are additional; under this identification the two recursor rules of CWS are the root rules over $\Sigma$.
\end{theorem}

\noindent Two schema isomorphisms agree on every element generated by the schema operations. With an empty set of variables, every free term is generated, so the isomorphism from the free schema onto the signature schema without additional symbols is unique; with variables, every permutation of the variables gives another isomorphism. On CWS the generated terms are those without $\mathrm{integrate}$, $\mathrm{merge}$, and $\mathrm{eqW}$, all isomorphisms agree there, and exchanging $\mathrm{merge}$ with $\mathrm{eqW}$ gives a second isomorphism that differs at $\mathrm{merge}(\mathrm{void},\mathrm{void})$.

\subsection{Every duplication arity}

\begin{theorem}[Coupling at every arity]\label{thm:arity-coupling}
  For $r\ge0$, let the $r$-ary rule $\mathrm{recur}(b,s,\mathrm{succ}(n))\to\mathrm{wrap}_r(s,\dots,s;\mathrm{recur}(b,s,n))$ emit $r$ copies of the payload beside the recursive value. A measure orients the $r$-ary rule if and only if its wrapper cost lies below its counter gain at every triple. If $M$ retains every wrapper argument at cost $c_w$ and orients the rule, the gain exceeds $r\,M(s)+c_w$ at every triple. For $r\ge1$, a bounded gain excludes orientation when the payload values are unbounded, and also when the wrapper cost is unbounded. At $r=0$ a retentive measure with bounded gain and unbounded values orients the rule, so the barrier uses $r\ge1$. At $r=1$ the barrier is Theorem~\ref{thm:coupling-inequality}, and the two statements are equivalent propositions.
\end{theorem}

\subsection{System~T}

G\"odel's System~T extends the simply typed $\lambda$-calculus with natural numbers, an iterator $\mathrm{It}$, and a recursor $\mathrm{R}$. We work with its raw syntax: free and bound variables, $0$, $S$, application, abstraction, $\mathrm{It}$, and $\mathrm{R}$. One step of reduction is beta, one of the rules
\[
  \mathrm{It}\,b\,s\,0\to b,\quad
  \mathrm{It}\,b\,s\,(S\,n)\to s\,(\mathrm{It}\,b\,s\,n),\quad
  \mathrm{R}\,b\,s\,0\to b,\quad
  \mathrm{R}\,b\,s\,(S\,n)\to s\,n\,(\mathrm{R}\,b\,s\,n),
\]
or a step inside any constructor. The iterator rule is the free duplicating rule with application as the wrapper. The recursor rule copies the step argument and the counter.

\begin{theorem}[The iterator is the free recursor]\label{thm:system-t-iterator}
  Let $\phi$ map free recursor syntax to System~T terms by $\mathrm{zero}\mapsto0$, $\mathrm{succ}\mapsto S$, $\mathrm{wrap}(s,y)\mapsto s\,y$, $\mathrm{recur}(b,s,n)\mapsto\mathrm{It}\,b\,s\,n$, and variables to free variables.
  \begin{enumerate}
    \item $\phi$ is injective, and its image is the set of terms without abstraction, bound variables, and recursor.
    \item System~T reduction from a term in the image stays in the image, and $t\Rightarrow_c u$ holds if and only if $\phi(t)$ reduces to $\phi(u)$ in one System~T step. Hence $\phi$ is an isomorphism of rewrite relations from free recursor syntax onto its image.
    \item The image, with the operations $0$, $S$, application, and $\mathrm{It}$, is a step-duplicating schema isomorphic to the free schema. Theorems~\ref{thm:orientation-criterion}, \ref{thm:coupling-inequality}, and~\ref{thm:four-cells} hold on it, and every term in the image is strongly normalizing for System~T reduction.
    \item A map of schemas from the free schema to System~T with the iterator reading is determined by its values on variables, and every assignment of terms to variables determines one. Sending a variable to a beta redex gives a map of schemas that fails to reflect steps.
  \end{enumerate}
\end{theorem}

\noindent Item (iv) states the other maps of the same kind: the maps of schemas form a set in bijection with the variable assignments, and $\phi$ is the assignment of free variables to free variables. Strong normalization in (iii) is the free contextual termination theorem transported along the isomorphism.

\begin{theorem}[The recursor is a three-place duplication]\label{thm:system-t-recursor}
  \begin{enumerate}
    \item No context $W$ with two holes satisfies $W[s,\mathrm{R}\,b\,s\,n]=s\,n\,(\mathrm{R}\,b\,s\,n)$ for all terms $b,s,n$. The three-place wrapper $\mathrm{wrap}_3(s,n,y)=s\,n\,y$ realizes the recursor rule, and every two-place schema is the three-place schema whose wrapper ignores the counter.
    \item Let a measure $M$ retain all three arguments of a three-place wrapper at cost $c_w$ and orient its rule. Then $M(s)+M(n)+c_w<\mathrm{gain}_M(b,s,n)$ at every triple. Hence a counter gain bounded by a function of $b$ and $s$ excludes orientation when counter values are unbounded, and a gain bounded by a constant excludes it when payload values are unbounded.
    \item Give every variable and $0$ weight one, let $S$ add one and application add one to the sum of its arguments, and read both $\mathrm{It}\,b\,s\,n$ and $\mathrm{R}\,b\,s\,n$ as $(n+1)(s+b+2)$. This interpretation orients every iterator instance, orients a recursor instance if and only if the counter weighs less than the base, and fails to orient the recursor rule. Adding $n^2$ to the recursor clause gives an interpretation that orients both rule shapes, and every System~T term without abstraction is strongly normalizing for System~T reduction.
    \item Over simply typed applicative terms, with the recursor as a typed function symbol, the clauses of the higher-order recursive path order and of the computability path order accept the recursor rules for every precedence and status, and so do the computable closure of the general schema and the size-decrease check of sized types. Under the side conditions of each method, acceptance yields termination of recursor rewriting in this presentation through an embedding into a recursive path order.
  \end{enumerate}
\end{theorem}

\noindent The two-place schemas embed into the three-place schemas as those whose wrapper ignores the counter, and item (i) shows that the recursor of System~T lies outside that embedding for every context wrapper. Item (iii) measures the difference: the free polynomial escape has a counter gain independent of the counter, which the three-place coupling inequality excludes, while the gain of the squared-counter interpretation grows with the counter.

\subsection{Transport laws and cost accounting}

The transfers of this paper use the preservation laws below, and each law has a premise that its use needs. Table~\ref{tab:transport-laws} lists each transfer with its premise and a control example that fails when the premise is removed.

\begin{proposition}[Transport laws]\label{prop:transport-laws}
  Let $R\subseteq S$ be relations on one set and let a map $\phi$ send a relation $R$ on $A$ to a relation $R'$ on $B$.
  \begin{enumerate}
    \item If no measure in a class orients $R$, then no measure in the class orients $S$; in particular a root barrier is a contextual barrier.
    \item If $\phi$ is a bijection and $R'$ is the renamed relation, then $R'$ terminates if and only if $R$ terminates.
    \item If every $R$-step maps to a nonempty $R'$-path, then termination of $R'$ implies termination of $R$, with or without injectivity of $\phi$. A simulation that sends a step to a path of length zero transfers nothing: a looping one-state relation maps to the empty relation.
    \item If a relation terminates on the classes of an equivalence, it terminates.
    \item If $\phi$ is injective, reflects steps, and has an image closed under outgoing $R'$-steps, then confluence of $R$ implies confluence of $R'$ from every image point.
  \end{enumerate}
\end{proposition}

\begin{table}[ht]
  \centering\small
  \caption{Each transfer used in the paper holds under its stated premise, and the listed control fails without it.}
  \label{tab:transport-laws}
  \begin{tabularx}{\textwidth}{@{}P{3.1cm}P{3.6cm}Y@{}}
    \toprule
    Transfer & Premise & Control without the premise \\
    \midrule
    Root barrier to contextual barrier & Every root step is a contextual step & The minimum-wrapper interpretation of Theorem~\ref{thm:context-lift}(iii) orients both root rules and fails in a context \\
    Root orientation to contextual orientation & Strict context laws & The same minimum-wrapper interpretation \\
    Stability under substitution & Root and contextual steps are closed under substitution on free syntax & Proved for both relations; the controls of Section~\ref{sec:criterion} use closed terms only \\
    Barrier under added rules & The added rules keep the duplicating rule & Adding a self-loop keeps every retentive barrier and destroys termination \\
    Renaming of symbols & A bijection of carriers commuting with the operations & A fold into a one-point algebra identifies distinct constructors \\
    Termination from a target & Every source step becomes a nonempty target path & A constant encoding sends a looping source to the empty target \\
    Termination through a quotient & Termination of the class relation & An associative-commutative comparison that relates a term to itself \\
    Confluence along an embedding & Injectivity, step reflection, closure of the image & A collapsing fold equates every term \\
    Termination of a shared representation & Every shared step reads back to a nonempty source path & A shared step with the same read-back and a target loop \\
    \bottomrule
  \end{tabularx}
\end{table}

\noindent Every cost bound of the paper counts rewrite steps of one named relation, and the size of a term counts its constructors and variables. Theorem~\ref{thm:free-exponential} counts contextual steps of the free recursor and the call chains of its dependency-pair relation; Theorem~\ref{thm:dp-normalization} counts applications of the pair rule; Proposition~\ref{prop:full-context-complexity} counts contextual steps of CWS; Propositions~\ref{prop:safestep-complexity} and~\ref{prop:root-normalizer-envelope} count guarded contextual and guarded root steps. Serialization of terms, the checking of certificates, and the arithmetic inside an interpretation are outside these counts. Duplication in a tree copies the step argument, while a shared representation keeps one copy, and the Lean development classifies the corresponding shared-representation cases.

\section{The CWS case study}
  \label{sec:certification}
  The free recursor isolates the orientation boundary. CWS adds six surrounding rules and a guarded variant, giving a fixed system on which termination, confluence, normalization, and complexity can be compared with the direct barriers. We write $R_g$ for the guarded root relation, $R_g^{\mathrm{ctx}}$ for its partial context closure, and $R$ for the unguarded root relation.

  \subsection{Guarded termination and confluence}\label{sec:sn}
  The root termination proof uses the lexicographic measure
  \[
    \mu(t)=(\delta\text{-flag}(t),\kappa^M(t),\tau(t))
  \]
  over $\mathbb N\times\mathrm{Multiset}(\mathbb N)\times\mathbb N$. The first component marks a recursor whose counter begins with $\delta$; the multiset records recursive-call weights; and $\tau$ is a weighted constructor size.

  \begin{theorem}[Per-step decrease]\label{thm:stepdrop}
    Every guarded root step $t\to_g t'$ satisfies $\mu(t')<_{\mathrm{lex}}\mu(t)$. Hence $R_g$ is strongly normalizing.
  \end{theorem}
  \begin{proof}
    On the successor recursor rule the first component drops from $1$ to $0$. On rec-zero and guarded merge cancellation the multiset component decreases in the Dershowitz-Manna order. In every remaining rule the first two components tie and $\tau$ decreases.
  \end{proof}

  \begin{theorem}[Guarded confluence and unique normal forms]\label{thm:confluence}
    The guarded root relation $R_g$ is confluent. Every term therefore has a unique guarded root normal form.
  \end{theorem}
  \begin{proof}
    The rule guards remove the only unjoinable equality overlap and reduce the remaining root peaks to a finite joinable set. Strong normalization from Theorem~\ref{thm:stepdrop} then gives confluence by Newman's lemma.
  \end{proof}

  The same guarded rules remain terminating and confluent under the partial context closure used in the case study. The context proof uses a strictly decreasing numeric potential whose $\delta$ clause is exponential and whose recursor clause weights the local counter and payload enough to dominate the duplicated branch.

  \subsection{The unguarded context relation}

  \begin{theorem}[Full context-closed strong normalization]\label{thm:full-context-closed-sn}
    The full contextual closure of the unguarded CWS root relation is strongly normalizing.
  \end{theorem}
  \begin{proof}
    The nonlinear polynomial interpretation $W$ from Proposition~\ref{prop:poly-full-step} is strictly monotone in every constructor argument and strictly decreases on each root rule. Hence it decreases under every context.
  \end{proof}

  \begin{theorem}[Full-context nonconfluence]\label{thm:full-context-nonconfluence}
    The full contextual CWS relation is neither locally confluent nor confluent. In particular,
    \[
      \mathrm{eqW}(\mathrm{void},\mathrm{void})
    \]
    reduces to the two distinct contextual normal forms $\mathrm{void}$ and $\mathrm{integrate}(\mathrm{void})$.
  \end{theorem}
  \begin{proof}
    The reflexive equality rule gives $\mathrm{void}$. The unequal-case rule gives $\mathrm{integrate}(\mathrm{merge}(\mathrm{void},\mathrm{void}))$, and a merge contraction under the integration context gives $\mathrm{integrate}(\mathrm{void})$. Neither target reduces further.
  \end{proof}

  Thus termination holds without the equality guard, and confluence fails without it. This separates the termination certificate from the guard conditions used by the confluence proof.

  \subsection{Normalization and complexity}\label{sec:normalizer}
  Because $R_g$ is terminating and deterministic at the root, well-founded recursion defines its normalizer. Confluence makes the normal form unique.

  \begin{proposition}[Linear guarded root-normalizer envelope]\label{prop:root-normalizer-envelope}
    Let $N_g(t)$ be the number of guarded root contractions performed by the deterministic normalizer. Then
    \[
      N_g(t)+2\le 2|t|.
    \]
    The merge-spine family $m_0=\mathrm{void}$, $m_{n+1}=\mathrm{merge}(\mathrm{void},m_n)$ requires $n$ root contractions, so the linear scale is attained.
  \end{proposition}

  \begin{proposition}[Single-exponential guarded contextual complexity]\label{prop:safestep-complexity}
    Every $R_g^{\mathrm{ctx}}$ derivation from a term of structural size $n$ has length $m$ satisfying
    \[
      m+1\le 2^{2n}.
    \]
    There is also an explicit family of size $5n+3$ admitting at least $2^n$ guarded contextual contractions. Hence guarded contextual derivational complexity is single-exponential up to constant factors in the exponent.
  \end{proposition}

  \begin{proposition}[Single-exponential full-context complexity]\label{prop:full-context-complexity}
    If $t\to_R^{m}u$ is a full contextual unguarded reduction, then
    \[
      m+W(u)\le W(t)\le 8^{|t|},
    \]
    so $m<8^{|t|}$. In the other direction, there is a family of size $5n+3$ with reductions of length $2^{n+2}-3$. Thus the fixed full-context system has matching single-exponential upper and lower scales.
  \end{proposition}

  \paragraph{Further proof data.}
  The Lean development gives the guarded contextual potential, representative local joins, the multiset order type $\omega^\omega$, the two-block calibration $\omega^\omega\!\cdot\!2$, realized lower families, and the detailed complexity constructions. These refine the fixed-system case study.

\section{TTT2 and CeTA certification of the CWS context relation}
\label{sec:ttt2}

This section concerns one fixed external object: the eight-rule CWS TRS obtained by dropping the \texttt{SafeStep} guards and using ordinary context-closed rewriting. TTT2~\cite{ttt2} proves termination by three strategies, and CeTA validates the resulting certificates. These results corroborate Theorem~\ref{thm:full-context-closed-sn} for this one system. A TTT2 MAYBE result records a failed search by that strategy.

\paragraph{Result.} TTT2 proves termination of the context-closed TRS using the DP framework:
\begin{enumerate}
  \item Dependency pairs extracted: $\mathrm{rec}\Delta^\sharp(b,s,\delta(n))\to\mathrm{rec}\Delta^\sharp(b,s,n)$, plus two non-cyclic \texttt{eqW} pairs.
  \item SCC decomposition: singleton $\{\mathrm{rec}\Delta^\sharp\}$.
  \item Subterm criterion with $\pi(\mathrm{rec}\Delta^\sharp)=3$: third argument decreases from $\delta(n)$ to $n$.
\end{enumerate}

\begin{table}[htbp]
  \centering
  \scriptsize
  \setlength{\tabcolsep}{3pt}
  \caption{TTT2 strategy results. Strategy labels are the TTT2 presets FAST and COMP, polynomial interpretation (POLY), matrix interpretation (MAT), and forward/backward instantiation (FBI). CeTA column: independent certification by CeTA~2.36~\cite{CeTA236}.}
  \label{tab:ttt2-strategies}
  \begin{tabularx}{\textwidth}{@{}l>{\raggedright\arraybackslash}Xc c c@{}}
    \toprule
    Strategy & Method class & Result & Time & CeTA \\
    \midrule
    FAST & DP + SCC + subterm criterion & \textbf{YES} & 0.06s & \textsc{certified} \\
    COMP & DP + termination dependency graph (TDG) + SCC + subterm + matrix + LPO & \textbf{YES} & 0.70s & \textsc{certified} \\
    LPO & Lexicographic path order (LPO) & \textbf{YES} & 0.02s & \textsc{certified} \\
    \midrule
    POLY & Polynomial interpretation & MAYBE & 0.15s & no certificate \\
    KBO & Knuth-Bendix order & MAYBE & 0.02s & no certificate \\
    MAT(2) & Matrix interpretation dim 2 & MAYBE & 0.31s & no certificate \\
    MAT(3) & Matrix interpretation dim 3 & MAYBE & 0.37s & no certificate \\
    FBI & Forward/backward instantiation & MAYBE & 0.17s & n/a \\
    \bottomrule
  \end{tabularx}
\end{table}

\noindent The three YES rows use dependency-pair or path-order methods. The POLY, KBO, MAT(2), and MAT(3) MAYBE rows agree with the direct barrier theorems of Section~\ref{sec:barrier}, which prove impossibility for the stated direct families. The FBI row records an external search failure.

\noindent The KBO result is explained by Corollary~\ref{cor:kbo-impossible}: any comparator satisfying the variable condition rejects the duplicating orientation at the rule schema. The POLY row is covered for the direct polynomial families proved above, while the successful nonlinear interpretation $W$ lies outside them through counter-payload coupling and successor non-transparency. The natural-matrix theorem explains the MAT$(2)$ and MAT$(3)$ failures whenever both wrapper matrices have a positive diagonal entry at a tracked coordinate.

\noindent In the FAST certificate, the dependency graph has one recursive SCC, the third recursor argument is selected, and that argument decreases from $\delta(n)$ to $n$. CeTA checks the full generated certificate; the paper uses the independent certificate only as validation of the fixed-system result.

\paragraph{Independent certification.}
All three YES strategies received a \textsc{certified} decision from CeTA~2.36. The POLY, KBO, MAT(2), and MAT(3) strategies answered MAYBE and produced no certificate for CeTA to replay; FBI also answered MAYBE, and its CeTA column reads n/a. Each MAYBE records a failed search by a bounded heuristic strategy. These external tool runs corroborate the method separation predicted by the barrier theorems.

\paragraph{Free recursor export.}
The exporter also writes the two-rule free recursor as \path{Artifacts/ttt2/free_recursor.trs}. TTT2 1.19 FAST returned YES in $0.023$s. CeTA returned CERTIFIED. The CPF proof uses the dependency-pair framework and the subterm criterion on the third argument of \texttt{recur}, matching the Lean theorem \path{freeRecursorTRS_terminating}.

\section{Lean formalization and public artifact}
\label{sec:formalization}

Every numbered definition, proposition, theorem, and corollary of this paper is checked in Lean~4, and the development follows the mathematical decomposition of the paper. This section records the representation, the boundary as a formal object, the headline correspondence, the generic infrastructure the development contains, and the trust boundary. The complete correspondence, one row per numbered claim, is the file \texttt{1.ORIENTATION\_BOUNDARY\_CLAIM\_INDEX.md} of the public artifact.

\subsection{Representation}
A step-duplicating schema is a structure over an arbitrary carrier type with the four operations of Definition~\ref{def:schema}, and every schema theorem quantifies over it. The free recursor is an inductive type of terms with variables over an arbitrary type; its root and contextual relations are inductive relations on that type. The Concrete Witness System is a second inductive type of ground terms, with the relations \texttt{Step}, \texttt{SafeStep}, and \texttt{SafeStepCtx} of Table~\ref{tab:relations-overview} and the full context closure \texttt{StepCtxFull}. The scalar grammar of Theorem~\ref{thm:grammar-closure-literal} is an inductive type of expressions with an evaluation function on the $(\mathrm{counter},\mathrm{payload})$ carrier, and the observations of Theorem~\ref{thm:free-profile-attainment} are recursive functions on free terms, so Theorem~\ref{thm:orientation-boundary-actual} is stated on actual terms and actual root steps. Decision procedures are executable functions with soundness and completeness theorems, and a rejected input returns its failing triple as a value. Each of the seventy-six method rows is a record of typed method data, the laws of the method, and its result, and the classification theorem is a dependent function over the inductive type of rows. The development declares no axiom of its own.

\subsection{The boundary as an object}
The Orientation Boundary is a set. Let $\mathcal O$ be the set of grammar expressions $e$ such that $M_e$ decreases across every actual root successor step of the free recursor, and let $\mathcal B$ be the set of expressions whose denotation is payload-independent and counter-strict. Theorem~\ref{thm:orientation-boundary-actual} is the equality $\mathcal O=\mathcal B$; the development proves the biconditional at every expression and records it as an equality of sets. Both sides of the boundary are inhabited: the counter projection lies in $\mathcal O$, and the payload projection and every constant lie outside it. The growth cells of Definition~\ref{def:growth-cells} form a four-element type; at every measure and pair one cell holds and no second cell does, and the barrier cell excludes orientation (Theorem~\ref{thm:four-cells}). The direct duplication law and the twelve family barriers are stated over the schema structure, so each applies to every instance: the textbook rule of Remark~\ref{rem:classic-instance}, System~T, and CWS. The countermodels of the twenty premise slots hold on the free schema (Theorem~\ref{thm:premise-necessity}). The seventy-six method rows are records over the free recursor.

\subsection{Headline correspondence}
Table~\ref{tab:headline-correspondence} lists the Lean declarations of the results a reader is likely to check first. Each declaration is stated over the carrier named in the paper theorem, and the public index gives the remaining rows.

\begin{table}[ht]
  \centering\small
  \caption{Headline results and their Lean declarations. Every module path is relative to the root of the Lean development in the public artifact.}
  \label{tab:headline-correspondence}
  \begin{tabularx}{\textwidth}{@{}P{2.1cm}YP{4.9cm}@{}}
    \toprule
    Result & Lean declaration & Module \\
    \midrule
    Thm.~\ref{thm:orientation-boundary-actual} & \path{free_successor_orientation_iff_payloadBlind_and_counterStrict} & \path{Meta/Methods/OrientationClosure/OrientationBoundaryPublicationCapstone.lean} \\
    The boundary as a set; the growth cells as a type & \path{orientationBoundary_eq}, \path{growthCell_eq_iff}, \path{barrier_cell_excludes} & \path{Meta/Methods/OrientationClosure/OrientationBoundaryObject.lean} \\
    Thm.~\ref{thm:free-profile-attainment} & \path{every_successor_profile_attained}, \path{derivationHeight_pointwise_least} & \path{Meta/OperationalInexpressibility/FreeRecursorWitnessOrder.lean} \\
    Thm.~\ref{thm:orientation-criterion} & \path{orients_all_iff_wrapperCost_lt_counterGain} & \path{Meta/Methods/OrientationClosure/CouplingTheorem.lean} \\
    Thm.~\ref{thm:four-cells} & \path{four_cell_coverage}, \path{barrier_cell_excludes_orientation}, \path{nonbarrier_cells_have_both_verdicts} & \path{Meta/Methods/OrientationClosure/CellClassification.lean} \\
    Thm.~\ref{thm:direct-duplication-law} & \path{no_unconstrainedDirect_orients_dup_step} & \path{Meta/NonlinearUnconstrainedExactLaw.lean} \\
    Thm.~\ref{thm:natural-matrix-barrier} & \path{no_natMatrix_orients_dup_step_of_tracked_strict} & \path{Meta/MatrixBarrierNatural_Schema.lean} \\
    Thm.~\ref{thm:tropical-context-barrier} & \path{no_tropical_context_orientation}, \path{tropical_root_context_classification} & \path{Meta/MatrixBarrierTropicalContextClosure_Schema.lean} \\
    Thm.~\ref{thm:subterm-coefficient-kbo} & \path{no_subtermCoefficient_variable_condition_orients_dup_step} & \path{Meta/SymbolicComparatorBarrier_Weighted_Schema.lean} \\
    Thm.~\ref{thm:premise-necessity} & \path{used_hypothesis_has_countermodel}, \path{slot_premise_exactly_necessary} & \path{Meta/Methods/OrientationClosure/DirectFamilyHypothesisNecessity.lean} \\
    Thm.~\ref{thm:escape-trichotomy} & the escape-trichotomy theorems for the scalar, tracked-vector, and projection universes & \path{Meta/EscapeTrichotomy.lean} \\
    Thm.~\ref{thm:method-classification} & \path{method_universe_closure}; \path{rowNecessityEvidence}, \path{rowNecessityKind_counts} & \path{Meta/Methods/OrientationClosure/MethodUniverseClosure.lean}; \path{Meta/Methods/OrientationClosure/MethodHypothesisNecessityCatalog.lean} \\
    Thm.~\ref{thm:tdec1} & \path{decideRootUnit_iff}, \path{failSuccUnit_spec} & \path{Meta/Methods/OrientationClosure/PolynomialOrientationDecision.lean} \\
    Cor.~\ref{cor:residual-zero-reduction} & \path{residual_decider_implies_bivariate_zero_decider} & \path{Meta/Methods/OrientationClosure/SeparableSlopesDecision.lean} \\
    Thm.~\ref{thm:s10-arbitrary-semantic-classifier} & \path{orienting_measure_counter_dominated_of_payload_erasure} & \path{Meta/RDRSSemanticArbitraryClassifier.lean} \\
    Thm.~\ref{thm:free-source-normalization} & \path{freeRecursor_ctx_rev_wellFounded}, \path{freeRecursor_ctx_joinable_iff} & \path{Meta/OperationalInexpressibility/FreeRecursorContext.lean} \\
    Thm.~\ref{thm:free-exponential} & \path{exponential_sandwich} & \path{Meta/Methods/OrientationClosure/FreeDerivationalComplexity.lean} \\
    Thm.~\ref{thm:free-recursor-sorted-persistence} & \path{freeRecursor_sorted_persistence_iff} & \path{Meta/Methods/OrientationClosure/MethodRowsSortedConstrained.lean} \\
    Thm.~\ref{thm:system-t-iterator} & \path{iterRelIso}, \path{iterImageIso} & \path{Meta/Methods/OrientationClosure/SystemTRecursor.lean} \\
    Thm.~\ref{thm:full-context-closed-sn}, Thm.~\ref{thm:confluence} & \path{wf_StepCtxFullRev_poly}; \path{confluentSafe} & \path{Meta/ContextClosed_SN_Full.lean}; \path{Meta/Newman_Safe.lean} \\
    \bottomrule
  \end{tabularx}
\end{table}

\subsection{Generic infrastructure}
The recursive path order results, the higher-order fragment, and the first two dependency-pair results below are stated for arbitrary signatures and rule sets and make no reference to the recursor; the processor-sequence result and many-sorted persistence are stated on the free recursor. The recursive path order with status is well founded for every well-founded precedence over every signature, by nested accessibility and without Kruskal's theorem, and every system whose rules it orients terminates under full contextual rewriting; the same two statements hold for the order modulo permutative congruence, and the associative-commutative variant obtained by flattening is well founded. A fragment of the higher-order recursive path order on applicative typed terms is sound through an embedding into the first-order order. In the dependency-pair framework, finiteness of the dependency-pair problem of a rule set satisfying the variable condition gives termination of that set, the reduction-triple processor is sound, and on the free recursor a sequence of sound processors is sound. Many-sorted persistence is an equivalence for every sort assignment that types the free-recursor rules (Theorem~\ref{thm:free-recursor-sorted-persistence}). The escape rows of Section~\ref{sec:method-classification} use these theorems, which are stated in a form other developments can import.

\subsection{Trust boundary}
The toolchain is pinned to Lean~4 release \texttt{v4.22.0-rc4} and to one Mathlib commit. The proof closure of every theorem in this paper uses only the three standard axioms of Lean's foundation, propositional extensionality, quotient soundness, and choice; no declaration uses \texttt{sorry}, and the check files under \texttt{Test/} print the axioms of every declaration cited in the paper. The public root module imports every module of the artifact, so one build command kernel-checks the whole release, and the reproducibility executable re-checks the generated TPDB exports against the certified files. TTT2 and CeTA act on the exported systems only.

\section{Related work}
\label{sec:related}

\paragraph{Termination of duplicating systems.}
The multiset ordering of Dershowitz and Manna~\cite{DershowitzManna79} handles rules replacing one term by multiple smaller terms. Recursive path orderings~\cite{Dershowitz87} extend this to simplification orders. Duplication and termination are surveyed in~\cite{Terese03,BaaderNipkow98}. Zantema~\cite{Zantema01} established a hierarchy between termination method classes. Endrullis, Waldmann, and Zantema~\cite{EndrullisWaldmannZantema08} developed matrix interpretations as a positive termination method; Theorem~\ref{thm:natural-matrix-barrier} shows that they fail on the duplicating rule whenever both wrapper matrices have a positive diagonal entry at a tracked coordinate.

\paragraph{Interpretation frameworks.}
Yamada's tuple interpretations~\cite{YamadaTuple22} give a taxonomy of polynomial, matrix, arctic, and related interpretation methods. We use that work to place the direct fragments studied here. The barriers for the twelve direct classes are proved for the scalar and tracked vector and pair interfaces stated in the theorems. The additive-sum Yamada fragment is a barrier by reduction to the additive theorem (Proposition~\ref{prop:yamada-additive-sum}), and no interpretation of the affine strict-coordinate fragment with positive wrapper coefficients orients the duplicating step (the same proposition). The successful escape mechanisms in this paper use transformed relations (dependency pairs) or global examples outside the blocked direct fragments.

\paragraph{Dependency pairs and modular termination.}
The dependency pair method~\cite{ArtsGiesl00} was introduced to overcome limitations of direct simplification orders, and Hirokawa and Middeldorp~\cite{HirokawaMiddeldorp04} introduced the subterm criterion. Thiemann and Giesl showed that size-change termination for TRSs captures only simple termination, while dependency pairs also handle systems that are not simply terminating~\cite{ThiemannGiesl05}. IsaFoR/CeTA~\cite{IsaFoR}, CoLoR~\cite{CoLoR}, and Coccinelle/CiME~\cite{ContejeanEtAl07} mechanize or certify positive termination arguments. The present work uses these systems as comparison points while proving a different kind of result: impossibility theorems for direct method classes on a fixed terminating schema.

\begin{table}[ht]
  \centering
  \small
  \caption{Position of this work relative to nearby formal developments.}
  \label{tab:related-comparison}
  \begin{tabular}{@{}P{0.20\linewidth}P{0.13\linewidth}P{0.23\linewidth}P{0.16\linewidth}P{0.16\linewidth}@{}}
    \toprule
    Work & Proof assistant & Primary focus & Fixed-system impossibility theorems & External certificate checking \\
    \midrule
    This work & Lean~4 & Barrier theorems over every schema, escape classification, premise countermodels, generic path-order and dependency-pair soundness, and the CWS case study & Yes & Fixed-system certificate validation \\
    AFP undecidability formalization~\cite{MitterwallnerMiddeldorpThiemannAFP24} & Isabelle/HOL & Undecidability of orientation problems across all inputs & Absent & Absent \\
    IsaFoR/CeTA~\cite{IsaFoR} & Isabelle/HOL & Certification and checking of positive TRS proofs & Absent & Present \\
    CoLoR / Coccinelle-CiME~\cite{CoLoR,ContejeanEtAl07} & Coq & Certified checking / soundness for positive rewriting arguments & Absent & Present \\
    Lean Metatheory~\cite{RamosMetatheory25} & Lean~4 & Generic rewriting, confluence, and normalization infrastructure & Absent & Absent \\
    \bottomrule
  \end{tabular}
\end{table}

\paragraph{Path orders and proof-theoretic strength.}
Ferreira and Zantema~\cite{FerreiraZantema95} prove a well-foundedness criterion from the subterm property together with decomposability, with completeness results over finite alphabets. Buchholz~\cite{Buchholz95} analyzes LPO termination and derivation-length bounds, while Rathjen and Weiermann~\cite{RathjenWeiermann93} calibrate Kruskal's theorem. These general results give proof-theoretic context for path orders. The CWS MPO proof here instead uses the concrete ordinal ranking of Proposition~\ref{prop:mpo-proof-theoretic-bound}. The DP subterm criterion~\cite{HirokawaMiddeldorp04} uses only the subterm relation, and its soundness is formalized in IsaFoR~\cite{SternagelThiemann10}.

\paragraph{Confluence and mechanization platforms.}
Newman's lemma~\cite{Newman42} has been mechanized in Isabelle/HOL~\cite{SternagelThiemannAFP10} and Lean~4~\cite{RamosMetatheory25}. Our use is classical: strong normalization plus the explicit local-join analysis gives confluence of the guarded relation. IsaFoR/CeTA, CoLoR, and related systems certify properties of particular rewrite systems; the present paper complements that work with theorems that bound classes of direct orientation arguments.
The unique-normal-forms paper of Rahnama~\cite{rahnamaDistinction} gives a categorical view of the same contrast. In the present first-order setting, the recursor rule duplicates the step argument, while equality matching identifies repeated-variable occurrences. Linear logic records duplication as contraction~\cite{Girard1987}; Markov-category structure supplies copying and discarding~\cite{Fritz2020}.

\paragraph{Proof-assistant termination checkers.}
Proof-assistant termination checkers such as Agda's Foetus discipline, Coq's guard condition, and Lean's structural termination checker use language-specific information outside the direct measure classes studied here. For the textbook duplicating rule $f(x,s(y))\to g(x,f(x,y))$, Proposition~\ref{prop:aggregation-projection} isolates the relevant restriction: a single grammar measure must be payload-independent and counter-strict; lexicographic comparison needs a payload-independent first coordinate; strict componentwise comparison needs every coordinate payload-independent. Successful checkers may therefore rely on transformed calls, recursive structure, typing, or other information outside these aggregations.

\section{Conclusion}
\label{sec:conclusion}

The opening question was which direct measures prove termination of a rule that copies its step argument while it consumes one recursion counter. For scalar grammar measures the answer is one set with two descriptions. Described by the rewrite steps, the orienters are the expressions that decrease across every actual root successor step of the free recursor; described by the expressions alone, they are the payload-independent, counter-strict expressions. Theorem~\ref{thm:orientation-boundary-actual} identifies the two sets, and every profile the classification uses is attained by a term.

The mechanism is the imbalance between wrapper cost and counter gain. A natural-valued measure that keeps both wrapper arguments and gains a bounded amount per counter step fails, the four-cell theorem shows that the barrier cell alone decides failure, and the premise-necessity theorem shows that every consumed premise of the twelve family theorems is used. Successful proofs violate a barrier hypothesis or use information outside the direct classes, in identifiable ways: dependency-pair projection forgets the duplicated payload, nonlinear coupling breaks successor transparency, path orders import subterm structure, and sharing removes the independence of the copies. Tropical interpretations separate the root from its context closure: a one-dimensional interpretation orients every root instance, and no finite-dimensional one orients the rule under both wrapper congruences. Seventy-six termination methods from the literature receive a machine-checked termination verdict on the free recursor.

At unit successor slope, the decision theorem decides the orientation of the two free-recursor rules by constructor-local natural polynomial interpretations of arbitrary degree: the procedure is total and returns a failing assignment on rejection. At higher slopes its undecided remainder is a signed positivity problem that contains the two-variable Diophantine zero problem over $\mathbb N^2$, and six strata of it are decided.

On the free recursor, orientation separates from termination. The two-rule system is strongly normalizing and confluent under all contexts, its derivations are single-exponential in term size, its escapes are calibrated at $\omega$ and $\varphi_4(0)$, and its root derivation height is the least natural-valued root ranking, which the counter and payload observations do not determine. For a step that carries a static payload erasure, every orienting semantic measure is counter-dominated, so no decisive payload-sensitive certificate exists. The boundary, the growth cells, the premise slots, and the method rows are objects of the Lean development, and the claim-to-code table of the public artifact has one row of Lean declarations for every numbered claim of this paper.

\appendix
\appendixlayout
\section{The seventy-six termination methods}\label{sec:methods}
Section~\ref{sec:method-classification} groups the seventy-six method families by the mechanism that decides their termination verdict. Table~\ref{tab:methods} lists every row with its citation and termination verdict; restricted languages are marked as fragments.

  \begin{longtable}{@{}P{0.46\linewidth} P{0.48\linewidth}@{}}
    \caption{Classification of seventy-six termination-method families on the free recursor.}\label{tab:methods}\\
    \toprule
    Method & Result on the free recursor \\
    \midrule
    \endfirsthead
    \toprule
    Method & Result on the free recursor \\
    \midrule
    \endhead
    \multicolumn{2}{@{}p{\linewidth}@{}}{\textit{Barrier cell (Theorem~\ref{thm:four-cells}): the wrapper cost grows with the copied argument and the counter gain does not}}\\
    Knuth-Bendix order~\cite{KnuthBendix70,DickKalmusMartin90,LudwigWaldmann07} & barrier \\
    Knuth-Bendix order with subterm coefficients~\cite{LudwigWaldmann07,YamadaKusakariSakabe15} & barrier \\
    Knuth-Bendix order with argument status~\cite{Steinbach89} & barrier \\
    Transfinite Knuth-Bendix order~\cite{LudwigWaldmann07} & barrier \\
    Lambda-free higher-order Knuth-Bendix order~\cite{BeckerBlanchetteWaldmannWand17} & barrier \\
    AC-compatible Knuth-Bendix order~\cite{YamadaWinklerHirokawaMiddeldorp14} & universal variable-condition obstruction and acceptance equivalence \\
    Matrix interpretations over the naturals~\cite{EndrullisWaldmannZantema08} & barrier \\
    Triangular matrix interpretations~\cite{MoserSchnablWaldmann08} & barrier \\
    Linear interpretations over the rationals~\cite{Lucas05} & barrier \\
    Linear interpretations over the reals~\cite{Lucas05} & barrier \\
    Matrix interpretations over the reals~\cite{Lucas10} & barrier \\
    \midrule
    \multicolumn{2}{@{}p{\linewidth}@{}}{\textit{Order of the carrier: contextual masking by max and min, strict chains in a finite order}}\\
    Arctic interpretations~\cite{KoprowskiWaldmann09} & barrier \\
    Tropical interpretations, the min-plus dual of arctic interpretations~\cite{KoprowskiWaldmann09} & barrier \\
    Strictly monotone algebras on a finite carrier~\cite{Zantema94} & barrier \\
    \midrule
    \multicolumn{2}{@{}p{\linewidth}@{}}{\textit{Derivation length (Theorem~\ref{thm:free-exponential}): the method bounds derivations linearly}}\\
    Match bounds~\cite{GeserHofbauerWaldmann04} & barrier \\
    Match bounds for left-linear systems~\cite{GeserHofbauerWaldmannZantema07} & barrier \\
    \midrule
    \multicolumn{2}{@{}p{\linewidth}@{}}{\textit{A syntactic condition of the method: right-linearity, one-step simulation, linear use of resources}}\\
    Right-forward closures~\cite{Dershowitz81} & barrier \\
    String rewriting~\cite{BookOtto93} & barrier \\
    Cycle rewriting~\cite{ZantemaKonigBruggink14} & barrier \\
    Linear typing without exponentials~\cite{Girard1987} & fragment barrier \\
    Type introduction~\cite{Zantema94} & barrier \\
    \midrule
    \multicolumn{2}{@{}p{\linewidth}@{}}{\textit{Coupling the counter gain to the payload (Theorem~\ref{thm:context-lift})}}\\
    Higher-order tuple interpretations~\cite{KopVale21} & typed-image tuple descent and free contextual termination \\
    Polynomial interpretations with negative coefficients~\cite{HirokawaMiddeldorp04b} & escape \\
    Max-polynomial interpretations~\cite{FuhsEtAl08} & escape \\
    Polynomial interpretations of any degree~\cite{ContejeanMarcheTomasUrbain05} & escape \\
    Multilinear interpretations~\cite{ContejeanMarcheTomasUrbain05} & escape \\
    Tuple interpretations, additive-sum fragment~\cite{YamadaTuple22} & barrier \\
    Strictly monotone algebras over the naturals~\cite{YamadaTuple22} & escape \\
    Extended monotone algebras~\cite{EndrullisWaldmannZantema08} & escape \\
    Generalized Knuth-Bendix order~\cite{MiddeldorpZantema97,YamadaKusakariSakabe15} & arbitrary-TRS termination from an orienting generalized KBO \\
    Knuth-Bendix order over polynomial weights~\cite{YamadaKusakariSakabe15} & escape \\
    \midrule
    \multicolumn{2}{@{}p{\linewidth}@{}}{\textit{Projection to the counter (Theorem~\ref{thm:dp-escape})}}\\
    Subterm criterion~\cite{HirokawaMiddeldorp04} & escape \\
    Argument filtering~\cite{ArtsGiesl00,KusakariNakamuraToyama99} & escape \\
    Reduction-pair processor~\cite{ArtsGiesl00,KusakariNakamuraToyama99} & escape \\
    Reduction triples~\cite{KopVanRaamsdonk12} & escape \\
    Neutral processors of the dependency-pair framework~\cite{GieslThiemannSchneiderKamp05} & escape \\
    Processor sequences of the dependency-pair framework~\cite{GieslThiemannSchneiderKamp05} & escape \\
    Size-change termination~\cite{LeeJonesBenAmram01,ThiemannGiesl05} & escape \\
    Usable rules~\cite{HirokawaMiddeldorp07} & fragment escape \\
    Formative rules~\cite{FuhsKop14} & escape \\
    Context-sensitive dependency pairs~\cite{AlarconGutierrezLucas10} & escape \\
    Persistence under sorts~\cite{Zantema94} & sorted termination of the free recursor gives \newline all-term termination; ill-sorted terms included \\
    Order-sorted dependency pairs~\cite{LucasMeseguer08} & escape \\
    Two-dimensional dependency pairs for conditional systems~\cite{LucasMeseguer14} & escape \\
    Operational termination of conditional systems~\cite{LucasMarcheMeseguer05} & escape \\
    Integer term rewriting~\cite{FuhsGieslPluckerSchneiderKampFalke09} & escape \\
    Logically constrained term rewriting~\cite{KopNishida13} & escape \\
    Higher-order logically constrained rewriting~\cite{GuoKop24} & arbitrary-signature applicative constrained root rewriting \\
    Size-change abstraction of a recursive definition~\cite{LeeJonesBenAmram01} & arbitrary covered ranking abstraction for free contextual rewriting \\
    \midrule
    \multicolumn{2}{@{}p{\linewidth}@{}}{\textit{Structural recursion on the counter}}\\
    Guard condition of fixpoint definitions~\cite{Gimenez95} & fragment escape \\
    Safe recursion with normal and safe inputs~\cite{BellantoniCook92} & fragment escape \\
    Ramified recurrence~\cite{Leivant95} & fragment escape \\
    Sized types~\cite{BartheFradeGimenezPintoUustalu04} & fragment escape \\
    General schema with the computable closure~\cite{BlanquiJouannaudOkada02} & fragment escape \\
    \midrule
    \multicolumn{2}{@{}p{\linewidth}@{}}{\textit{Precedence, subterm and ordinal-descent orders}}\\
    Simplification orders, recursive path order~\cite{Dershowitz87} & arbitrary-signature RPO soundness for every oriented TRS \\
    Recursive path order with status~\cite{Dershowitz87} & escape \\
    Polynomial path order~\cite{AvanziniMoser13} & escape \\
    AC-compatible recursive path order~\cite{RubioNieuwenhuis95} & escape \\
    Slow-growing hierarchy bound~\cite{MoserWeiermann03} & escape \\
    Higher-order recursive path order~\cite{JouannaudRubio99,JouannaudRubio07} & fragment escape \\
    Computability path order~\cite{BlanquiJouannaudRubio15} & fragment escape \\
    Quasi-decreasingness~\cite{Ohlebusch02} & escape \\
    \midrule
    \multicolumn{2}{@{}p{\linewidth}@{}}{\textit{Labelling}}\\
    Semantic labelling~\cite{Zantema95} & escape \\
    Predictive labelling~\cite{HirokawaMiddeldorp06} & escape \\
    Root labelling~\cite{SternagelMiddeldorp08} & escape \\
    Self-labelling~\cite{MiddeldorpOhsakiZantema96} & escape \\
    Match bounds with raise consistency~\cite{GeserHofbauerWaldmann04} & escape \\
    \midrule
    \multicolumn{2}{@{}p{\linewidth}@{}}{\textit{Changing the representation}}\\
    Rewriting modulo equations~\cite{JouannaudKirchner86} & nonconservative \newline escape \\
    Term graph rewriting with sharing~\cite{Plump99} & nonconservative \newline escape \\
    Weighted type graphs~\cite{BrugginkKonigNolteZantema15} & relative termination equivalence for strict and weak disjoint-sum graph rules \\
    Generalized weighted type graphs~\cite{EndrullisOverbeek26} & termination under compatible graph gluing; incompatible extension refuted \\
    Translation into a terminating process calculus~\cite{DengSangiorgi06} & generic termination reflection and synchronous counter-process translation \\
    Lambda-mu calculus through a CPS translation~\cite{Parigot92,deGroote94} & explicit-substitution erasing head beta with CPS-image descent \\
    Quasi-interpretations with sharing~\cite{BonfanteMarionMoyen11} & read-back bound with a covering reduction order; unpaired termination refuted \\
    Well-founded coalgebras~\cite{JeanninKozenSilva17} & no infinite path in any well-founded powerset coalgebra \\
    Infinitary rewriting~\cite{KennawayKlopSleepDeVries95} & finite termination and an infinite strongly convergent reduction \\
    \bottomrule
  \end{longtable}

\section*{Artifact availability}
The Lean~4 development, the TTT2 and CeTA certificates, and the reproduction instructions are public at \url{https://github.com/MosesRahnama/The-Orientation-Boundary}. The file \texttt{1.ORIENTATION\_BOUNDARY\_CLAIM\_INDEX.md} in that repository is the claim-to-code table of this paper: one row per numbered claim, with the Lean declarations that prove it, their module, their status, and the hypotheses they carry, followed by the module map and the source-file table. Every declaration name and module path in the file resolves in the repository tree at generation time, and each row names the claim by its number in this paper and by its label; a row that carries a label alone indexes a result stated in the Lean development. Building the root module with the pinned toolchain checks every result; the repository \texttt{README.md} gives the commands.

\end{document}